\documentclass[sn-mathphys-num]{sn-jnl}
\usepackage{graphicx}
\usepackage{amsmath,amssymb,amsfonts}
\usepackage{amsthm}
\usepackage[title]{appendix}
\usepackage{enumitem}

\newlist{myreferences}{description}{1}
\setlist[myreferences]{labelsep=0em, leftmargin=0em, 
                        itemindent=0em, 
                        itemsep=1em, 
                        style=nextline, 
                        after=\hangindent=2em \hangafter=1} 

\begin{document}

\title[Article Title]{Algorithmic Forecasting of Extreme Heat Waves}

\author[1]{\fnm{Richard A.} \sur{Berk}\email{berkr@sas.upenn.edu}}
\author[2]{\fnm{Amy} \sur{Braverman}\email{amy.braverman@jpl.nasa.gov}}
\author[3]{\fnm{Arun Kumar} \sur{Kuchibhotla}\email{arunku@cmu.edu}}

\equalcont{This work was partially performed at the Jet Propulsion Laboratory, California Institute of Technology under contract with the National Aeronautics and Space Administration. Government sponsorship acknowledged. Dr. Braverman acquired and organized the data and wrote the section on the AIRS data. Dr. Kuchibhotla formulated the data generation structure talking account of spatial dependence with a cross-over research design. Dr. Berk was responsible for the data analysis. The paper was jointly written.}

\affil[1]{\orgdiv{Statistics and Data Science}, \orgname{University of Pennsylvania}, \orgaddress{\street{265 37th Street}, \city{Philadelphia}, \postcode{19104-1686}, \state{Pennsylvania}, \country{U.S.A.}}}

\affil[2]{\orgdiv{Jet Propulsion Laboratory}, \orgname{California Institute of Technology}, \orgaddress{\street{4800 Oak Grove Drive}, \city{Pasadena}, \postcode{91109-8099}, \state{California}, \country{U.S.A}}}

\affil[3]{\orgdiv{Statistics \& Data Science}, \orgname{Carnegie Mellon University}, \orgaddress{{Baker Hall 132}, \city{Pittsburgh}, \state{Pennsylania}, \postcode{15213}, \country{U.S.A}}}

\abstract{This paper provides some foundations for valid forecasting of rare and extreme heat waves through a better understanding of the similarities and differences between several consecutive hot days under normal circumstances and rare, extreme heat waves. We analyze AIRS data from the North American, Pacific Northwest and AIRS data from the Phoenix, Arizona region. A genetic algorithm is used to help determine the most promising predictors. Classification accuracy with supervised learning is excellent for the Pacific Northwest and is replicated for Phoenix. Conformal prediction sets are considered as a way to represent forecasting uncertainty. Complications caused by endogenous sampling are discussed.
}


\maketitle

\section{Introduction}\label{intro}

Extreme heat waves are relatively rare events. But of late, extreme heat waves seem to be increasing in frequency and intensity (Mann et al., 2018). Despite a growing literature (Marx et al., 2021), Schmidt (2024) argues that ``heat anomalies'' may mean that climate science has underestimated the speed at which disastrous global warming will materialize. Many of the consequences for public health and local ecosystems are already troubling (Barriopedro et al., 2023).  Some projected consequences are dire (Khatana et al., 2024). The mass media also has taken note. From the Economist, for example, ``Deadly heat is increasingly the norm, not an exception to it.'' (Economist, June 6, 2024). 

Existing studies seeking to understand extreme heat waves have so far produced mixed results (Mann et al., 2018; McKinnon and Simpson, 2022, Fischer et al., 2023; Zhang et al., 2024; Li et al., 2024). Part of the problem is that during the warm-weather months, sequences of hot days have long been ``normal.'' These spells have certain features of heat waves that some have linked to climate change (Bartusek, 2022; Dosio et al., 2028). But they arguably are not the rare and extreme heat waves about which there has been so much recent concern.

In this manuscript, the primary goal is to help built a foundation on which valid forecasts of rare and extreme heat waves can be made. The data are taken from the Level 3 product of NASA's Atmospheric Infrared Sounder (AIRS) instrument aboard the Aqua spacecraft (Aumann et al., 2003; Pagano et al., 2010). Grid cells are the study units. Observations from the June 2021 heat wave in the North American, Pacific Northwest (PNW) are the initial focus. Observations from the 2023 July heat wave in the greater Phoenix, Arizona area allow for a replication. For both sites, ``control group'' comparisons constructed from ``usual'' summer days are taken from other months or other years for which no credible claims of extreme temperature episodes have been made. 

We offer several contributions. 

\begin{enumerate}
\item
We assemble daily, remote sensing data over many months that include an observed extreme heat wave. In that setting, heat waves are rare events. We employ a within-grid-cell research design that when implemented for sets of chosen days, constructively enables the analysis of rare events.
\item
We use as a response variable gain scores of grid cell, surface temperatures to address perhaps the most salient heat wave concern. The gain scores hold constant temporally stable features differing across grid cells (e.g., grid cell elevation), which necessarily cannot be directly associated with the dynamic nature of heat waves. Gains score distributions for the heat wave setting and the control settings are compared. In contrast to much past work, no ``off the shelf'' statistical distributions are assumed.
 \item 
We also compare the distributions of precursors two weeks before the actual heat wave to distributions of precursors two weeks before ordinary summer days, serving as control settings. Differences between the distributions can be important prerequisites for accurate heat wave forecasts. 
 \item
We then apply a genetic algorithm separately to heat wave data and comparison data to construct two synthetic populations, one for an ideal type extreme wave and one for ideal type ordinary summer days. For each, supervised learning applied to the numeric gain scores provides the requisite fitness functions. Several predictors surface strongly in their ability to differentiate the two synthetic populations and consequently possess forecasting promise, consistent with structural differences between extreme heat waves and commonplace hot summer days.  The promise of those predictors is replicated when the response variable is one of two discrete categories: a subsequent heat wave or no subsequent heat wave. 
 \item
 Absent any data exploration, we seek to reproduce our results in the greater Phoenix area in order to evaluate the robustness of our research design and substantive findings. The two climate settings could hardly be more different, but the results are qualitatively rather similar.
 \end{enumerate}
 
 \vspace{.1in}

Section 2 addresses some methodological challenges raised by efforts to forecast extreme heat waves, typically characterized by temperatures that likely fall in the far right tail of temperature distributions; they represent rare events that commonly cause serious data sparsity problems. Section 3 provides a grounded discussion of the remotely sensed, AIRS, grid cell data to be used and in Section 4, the research designs that can work well with rare events like extreme heat waves are introduced. Substantial progress is made on data sparsity and on design-based adjustments to mitigate the impact of nuisance variables. Section 5 uses gain scores to examine the marginal, ground-level temperature distributions of an actual heat wave compared to a faux heat wave. The distributions are rather different. Section 6 examines whether the predictor distribution differ as well. They do. Section 7 introduces a genetic algorithm to construct a synthetic population of ideal-type, extreme heat wave grid cells, and a synthetic population of ideal-type, ordinary high temperature summer grid cells. Comparisons between the predictor values in these different populations enable a simple form of feature engineering for variable selection. Section 8 uses the selected predictors to examine the promise of forecasting a binary response variable denoting whether or not a grid cell experiences an extreme heat wave. Section 9 reports an effort to reproduce the Pacific Northwest results using an extreme heat wave in the greater Phoenix area. As one would expect, the results somewhat differ, but the overall comparability of the broad conclusions is encouraging. Section 10 returns to the structure of our research designs and the matter of endogenous sampling. We consider the degree to which our efforts to overcome the rare event problem have introduced difficulties in the reported results. We suggest that there is promise in a reformulation from minimizing loss to minimizing risk making subject-matter and policy considerations are fundamental. Section 11 offers a brief summary of the results and some overall conclusions. The findings are framed as highly provisional in part because many more heat waves in a variety of locations need to be studied.

\section{Some Methodological Challenges}\label{challenges}
In the pages ahead, we use the term ``heat wave'' to denote extreme heat waves that are relatively rare and arguably enabled by climate change. We use the term ``heat dome'' as well to capture some instructive climate science thinking (e.g., Jain at el., 2024). Within this framing, there can be many definitions that differ at least in important details (Perkins, 2015; Piticar et al., 2019). The definitions we use are provisional, constrained in part by available information in the AIRS data.

Almost regardless of how heat waves are defined, a major statistical challenge is data sparsity because extreme heat waves are relatively unusual. One consequence is that very accurate but trivial forecasts can be guaranteed. If in most remotely sensed grid cells most of the time, there is no heat wave, a forecast of no heat wave will almost always be correct without needing to bother with predictors. When predictors are available, they typically will fail to improve the fit or forecasts because both are nearly perfect already. A more subtle consequence is that the design space defined by the predictors will be empty in many regions, making it nearly impossible to estimate well many kinds of nonlinear functions. In other words, for many combinations of predictor values, there are no data. The fallback of linear statistical methods can be overmatched and miss very important heat wave characteristics. Some consequences of these and related difficulties are addressed later.
 
Current global climate models (GCMs) do not appear to offer much help. For the spatial scales at which global climate models function, heat waves are generally below the spatial scales that can be well resolved. Various forms of downscaling seem necessary in the form of regional climate models (RCMs), but these introduce a number of new challenges (Gebrechorkos et al., 2023; Lafferty and Sriver, 2023; Schneider et al., 2023). As Gettelman and Rood  (2016) write,  ``For these smaller scale processes, we often must use statistical treatments to match observations to functions that can be used to describe the behavior. Some processes can be represented by basic laws [of physics], other processes must (at the scale of a climate model) be represented by statistical relationships that are only as good as our observations.'' To this we add: and only as good as the statistical models themselves (Breiman, 2001b; Freedman, 2009). Moreover, capitalizing on ``basic laws'' depends on knowing the relevant physics in particular, complex settings, getting boundary conditions from a GCM that are approximately correct, and having access to significant computational resources. And even then, the spatial scale may be too coarse. The analyses to follow suggest that certain quasi-experimental research designs can help when used together with supervised learning.

There are also difficulties in how best to represent uncertainty. Because there is apparently no consensus model for extreme heat waves, there is no widely acceptable way to formulate a generative model for how the observations materialize. Without such a formulation, the usual uncertainty calculations risk being little more than hand waving. Gettelman and Rood (2016) explain that in climate science, uncertainty is caused by (1) model uncertainty, (2) scenario uncertainty and (3) initial condition uncertainty. In statistics and econometrics, the first is called ``model misspecification'' (Freedman, 2009). The second is closely related to various forms of stationarity in statistical time series analysis (Box et al., 2016). The third overlaps in statistics with the literature on errors in variables (Fuller, 1987). How these and other statistical and econometric formulations map to those in climate science is largely unresolved. Spatial dependence complicates matters further.

Confusion also can be introduced by different meanings of the term ``model'' (Berk et al., 2024). In the natural sciences, a model typically is intended to facilitate explanation and, often, causal inference. The focus is on the manner in which some empirical phenomenon works, and understanding is commonly the dominant goal. Such models necessarily introduce simplifications to further understanding.  There also are various other ways models systematically can be right or wrong. 

An algorithm is not a model. ``An algorithm is nothing more than a very precisely specified series of instructions for performing some concrete task'' (Kearns and Roth, 2019). The goal is to perform that task well without necessarily representing the underlying subject-matter mechanisms. Algorithms are neither right nor wrong; as a technical matter, there is no such thing as a misspecified algorithm. Either an algorithm performs in a satisfactory fashion or it does not. For example, a pattern recognition algorithm might be used to identify malignancies in lung X-rays and often can perform better at that task than a radiologist (Li et al., 2022). But, there is in the algorithm no necessary explanatory or causal content for how the malignant tissue originated and grew.\footnote
{
Recent developments in physics-informed algorithms muddies the differences between models and algorithms (Kerniadakis et al., 2021). Formulations from physics, such as certain physical invariants, are embedded in special neural network architectures. Ideally, fitting performance is improved.
} 

\section{The AIRS Remotely Sensed Data}\label{AIRS}

The Aqua spacecraft responsible for the AIRS data was launched into polar orbit at an altitude of 705 km on May 4, 2002. Acquiring and processing AIRS data began in September of that year. AIRS observes Earth and its atmosphere in 2378 infrared spectral channels. The initial spectral input is processed  and stored as Level 1 products. (Aumann et al., 2020). Level 2 products result when the Level 1 Products are converted into geophyscial variables such as atmospheric moisture content at various altitudes (Suskind et al., 2020; Thrastarson, 2021). Level 3 data products (Tian et al., 2020) are intended for users who do not require the higher spatial and temporal resolution of Level 2. In the analyses to follow, we use Level 3 data.

\begin{table}
 \footnotesize
    \centering
    \begin{tabular}{|l|l|}
        \hline
        Name &  Description \\
        \hline\hline
        $Latitude$ &  Latitude at the center if the grid cell. \\
        \hline
        $Longitude$ & Longitude at the center of the grid cell.  \\
        \hline
        $LandSeaMask$ & 1 = land, 0 = sea.\\
        \hline
        $Topography$ & Topography of the Earth in meters above the geoid. \\
        \hline
        $StdPressureLev$ &  Pressure levels of temperature and water vapor profiles \\ 
        & (we use only 1 -- 12).  The array order is from the surface.\\
        & upward.\\
        \hline
        $SurfAirTemp$ & Temperature of the atmosphere at the Earth’s surface in \\
         & degrees Kelvin \\
        \hline
        $TropHeight$ &  Height of the tropopause in meters above sea level.  \\     
         \hline        
        $Temperature$ &  Temperature of the atmosphere in degrees Kelvin (we \\
          & use only 1 -- 12). The array order is from the surface.\\
          & upward.\\
        \hline
        $\mathit{H2O\_MMR}$ & Water vapor mass-mixing ratio in grams/kilogram.\\
        \hline        
    \end{tabular}
    \caption{AIRS Level 3 variables used in this study and brief descriptions.}
    \label{tab:variables}
\end{table}

The available Level 3 variables are shown in Table~\ref{tab:variables}. Their names are ``official'' in that those are the names that come with the downloaded data. We later shorten some names and change some names to facilitate data processing convenience and the construction of approachable tables and figures. But the mapping from the names in Table~\ref{tab:variables} to names used in subsequent tables and figures should be clear.\footnote
{
In , the phrase ``Pressure levels of temperature and water vapor profiles'' means that atmospheric pressure is proxy measure for altitude for both temperature and the water vapor profile. The former is in degrees Kelvin. The latter is a lot like relative humidity at different altitudes but is the proportion of the total atmospheric profile that is water vapor at a given altitude. The total atmospheric profile is the proportional representation of different constituents of the atmosphere at a given altitude such as oxygen, carbon dioxide, nitrogen, and water vapor. 
}

For the Level 3 variables in Table~\ref{tab:variables}, a potentially useful, statistical thought-experiment for the data generation starts with a fixed set of grid cells defined in the AIRS data. These are the fixed study units. There is one set of measures per grid cell per day. The variables measured do not change over time although for the four random variables, the values realized when measured can change: ``$StdPressureLev$", ``$TropHeight$'', ``Temperature'', and ``$\mathit{H2O\_MMR}$''. The fixed variables are ``$Latitude$,'' ``$Longitude$,'' ``$LandSeaMask$,'' and ``$Topography$''. Based on media accounts and past research, particular days of interest were determined before looking at the AIRS data. These days too are treated as fixed in the analyses to follow; they are not random variables. For each grid cell and each day, the values of the random variables are treated as if they are realized from putative, joint probability distributions that characterize local atmospheric conditions. These values are not presumed to be realized in an IID fashion because of associations between the variables in their joint distributions. In addition, the joint probability distributions likely differ somewhat by grid cell and day. 

\section{Research Design}\label{Design}

Any research design meant to anticipate extreme heat waves must address the question ``compared to what?'' The obvious answer is compared to ``non-heat waves.'' We begin with the 2021 Pacific Northwest extreme heat wave that affected parts of the U.S. and Canada. During the last four days of June, the Pacific Northwest experienced what many observers called an extreme heat wave. Although there are debates over what exactly constitutes an extreme heat wave, this event seems to qualify from almost any informed perspective. With so many extant definitions (Perkins, 2015, Barriopedro et al., 2023), and nothing close to a consensus on best practices, we will provisionally operationalize an extreme heat wave as a \emph{single discrete, unusually high temperature event over several days under a slowly moving ``heat dome''} (Li et al., 2024). More details are provided shortly. Other formulations certainly are possible, and perhaps even preferable, depending on the policy and/or research questions asked.\footnote
{
The many possible definitions of a heat wave leads to some clumsy wording. The term ``true'' heat wave implies broad definitional agreement. The same applies to ``actual'' heat wave. The use of ``extreme'' heat wave can be quite vague unless ``extreme'' is defined. Even the term ``heat wave'' also can mean that we are using a widely accepted definition. To reduce stale repetition, we use all four terms at various times depending on context with the understanding that the operationalization just provided is what we intend to convey. 
}

\begin{figure}[htbp]
\begin{center}
\includegraphics[width=5in, angle=0, origin=c]{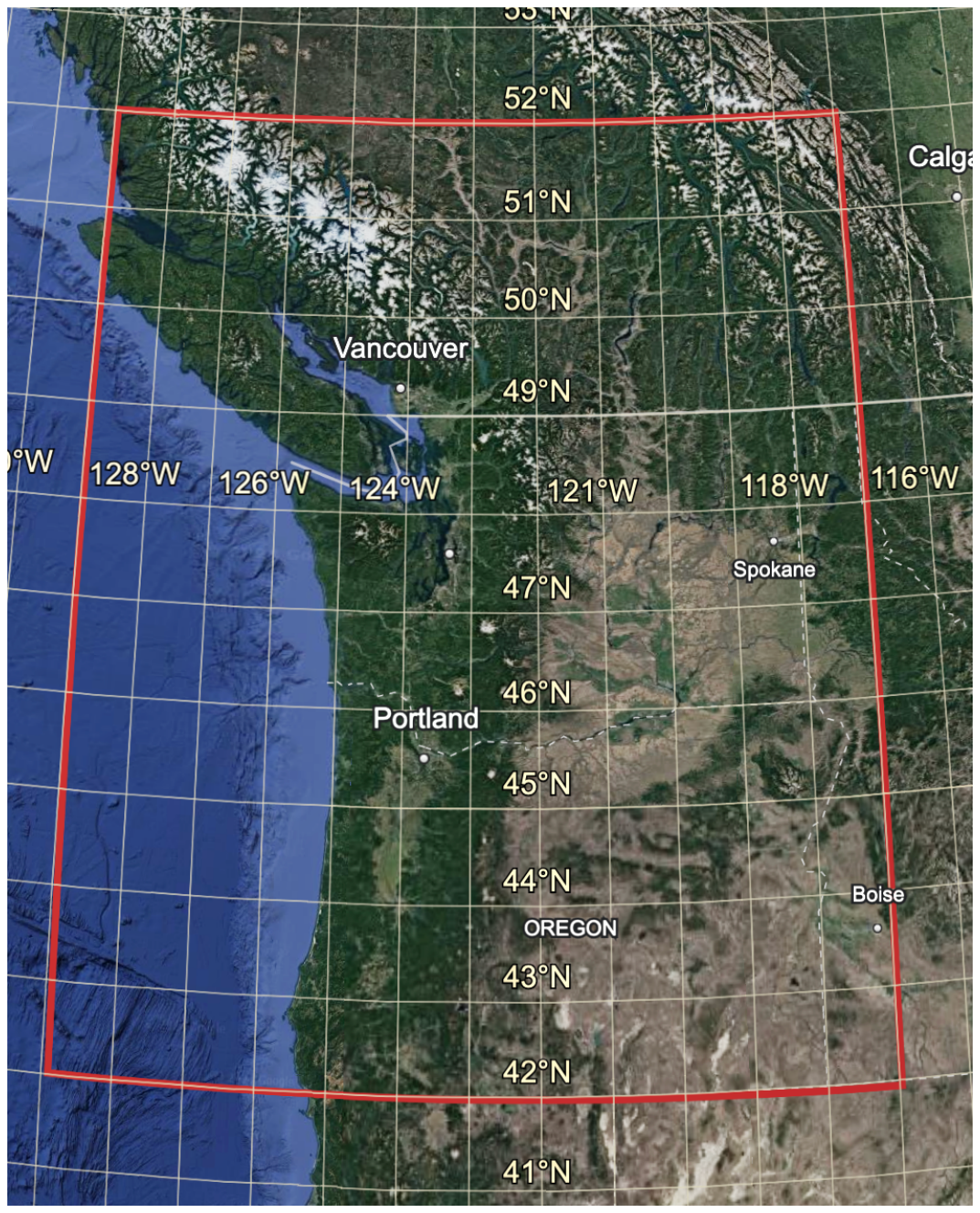}
\caption{U.S. Pacific Northwest study region shown by the large red rectangle. Smaller squares are grid cells. }
\label{fig:pnw_region}
\end{center}
\end{figure}

Media coverage of the June heat wave by and large was consistent with data from the Community Earth System Model (https://www.cesm.ucar.edu/), which is not employed in the current analysis. All relevant June days were confirmed with the PNW AIRS data used in the analyses to follow. The same sources were used to identify the absence of extreme heat waves in manner described shortly.  

Yet, the spatial boundaries of an extreme heat wave can be difficult to determine. One complication is a heat dome responsible can cover many square kilometers and move slowly over time. Different fractions of potentially affected grid cells, nominally in the same locale, will fall under the dome from day to day. In the case of the PNW June 2021 heat wave, the four-day definition seems to be a reasonable approximation.

Figure~\ref{fig:pnw_region} shows the location of our study area: a $12^\circ \times 12^\circ$ region over the northwest U.S. and southwestern Canada. For our analysis, we use daily data from these 144 grid cells, each grid cell indexed by location and day. The widely varying topography is apparent from Figure~\ref{fig:pnw_region}.

\begin{figure}[htbp]
\begin{center}
\includegraphics[width=5.6in]{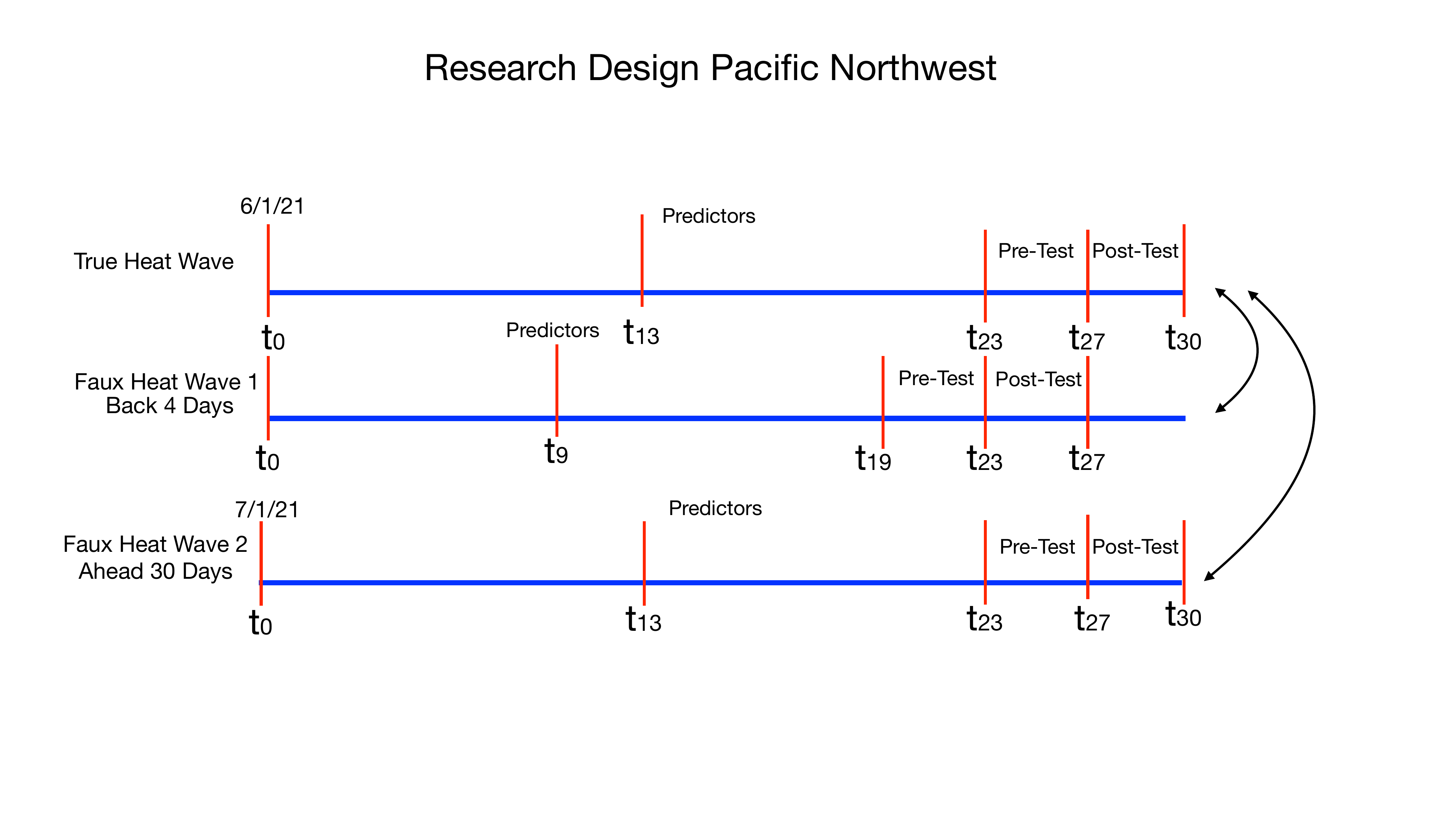}
\caption{The extreme heat wave data format at the top compared separately to Faux Heat Waves 1 or  2.}
\label{fig:Designs_PNW}
\end{center}
\end{figure}

To what should the PNW 2021 heat wave been compared?  The PNW AIRS curated data structure is shown in Figure~\ref{fig:Designs_PNW}: the top blue line for the extreme heat wave, the middle blue line for comparison ``Faux Heat Wave 1,'' and the bottom blue line for comparison ``Faux Heat Wave 2.'' The study units are the same 144 grid cells for the extreme heat wave and both faux heat waves; the same 144 grid cells are included within each of these settings. For variables not treated as fixed, the data values associated with a grid cell can differ depending on the day on which remotely sensed data were collected.

Starting with the heat wave at the top representing the month of June, 2021, the post-test is the mean surface temperature for \emph{each grid cell} over the 4 last days of the month. The pre-test is the mean surface temperature \emph{for each grid cell} over the 4 sequential days immediately preceding the first post-test day.  In both cases, means are used to help remove random variation in noisy measures of daily surface temperatures or variation too transient to be associated systematically with an extreme heat wave. The terms ``pre-test'' and ``post-test'' are common in experiments and observational studies seeking to estimate causal relationships. Our goal is to improve forecasting validity, which does not require estimates of cause and effect. Substantial associations between outcomes and outcome precursors can be sufficient. However, causal reasoning can often help to better formulate how the data are curated and organized. 

Having a pre-test as well as a post-test confers benefits discussed in the next section.  Based on past research (Li et al., 2034), all of the predictor values are taken from 14 days before the beginning of the true heat wave. We consider whether they are promising extreme heat wave precursors 2 weeks in advance. Other lag sizes are certainly possible and worth exploring. The same 144 grid cells constitute the post-test, pre-test, and lagged predictor observations.

Faux Heat Wave 1, shown as the middle blue line, has the same design configuration as the actual heat wave, but is shifted \emph{back} in time 4 days before the extreme heat began. Its post-test does not include the extreme heat wave, which provides one answer to the question ``compared to what?''  Faux Heat Wave 2 at the bottom also has the same design configuration as the extreme heat wave, but is shifted \emph{forward} in time one month to July, 2021, such that the extreme June heat wave also is not included in the post-test.  It provides as second answer to the question ``compared to what?'' For both faux heat waves, the same 144 grid cells constitute the post-test, pre-test, and lagged predictor  observations.

Faux Heat Wave 1 has the advantage of a pre-test falling immediately before the extreme heat wave post-test so that potential confounders might have quite similar values.  The risk is that some of the extreme heat wave characteristics may be in part a carried over from the immediately preceding several days; a larger temporal separation might be desirable (Benjamin-Chung et al., 2018). Therefore, Faux Heat Wave 2 provides a temporal separation of a month, but risks that summers in July may on the average differ from summers in June in ways confounded with surface temperatures.

In summary, the data generation thought experiment can be operationalized as Figure~\ref{fig:Designs_PNW}. Figure~\ref{fig:Designs_PNW} can be implemented using certain features of within-subject quasi-experimental designs (Maxwell et al., 2018). In particular, study units can be used as their own controls. We begin to capitalize on this in the next section.

\section{Temperature Changes from Pre-Test to Post-Test}\label{prepost}

We transition to empirical analyses by separately examining the PNW heat wave summer temperatures and a rendering of routine PNW summer temperatures. \emph{For each grid cell, the gain score is the post-test mean temperature minus the pre-test mean temperature.} This is just a routine difference in means computed separately for the actual heat wave and for each faux heat wave. Despite the name, gain scores can be positive or negative and are often called ``change scores;'' a gain score is the change from pre-test to post-test. For the analyses immediately to follow, attention centers on the change  between the 4-day pre-test mean surface temperature and the 4-day post-test mean surface temperature.
 
 We stress that although causal processes are involved, no causal effects are being explicitly estimated from the gain scores (Tennant et al., 2022). We are capitalizing on pre-test/post-test \emph{associations} to help simplify the analyses.  For example, some grid cells are routinely warmer or cooler than others because of features such as urban heat islands, elevation, type of ground cover, and proximity to the Pacific Ocean or the Cascade Mountains. On moderate time scales, these location features do not change. Therefore, they cannot by themselves be directly associated with extreme heat waves that are necessarily dynamic over a period of several days; a constant cannot be associated with a variable. We treat them as fixed location features controlled by the use of gain scores (Maris, 1998, Maxwell et al., 2018). Only temporally dynamic variables can directly matter.
 
 \subsection{Comparisons Between Gain Score Distributions for the Actual Heat Wave and Faux Heat Wave 1}\label{histo} 

\begin{figure}[htbp]
\begin{center}
\includegraphics[width=4.5in]{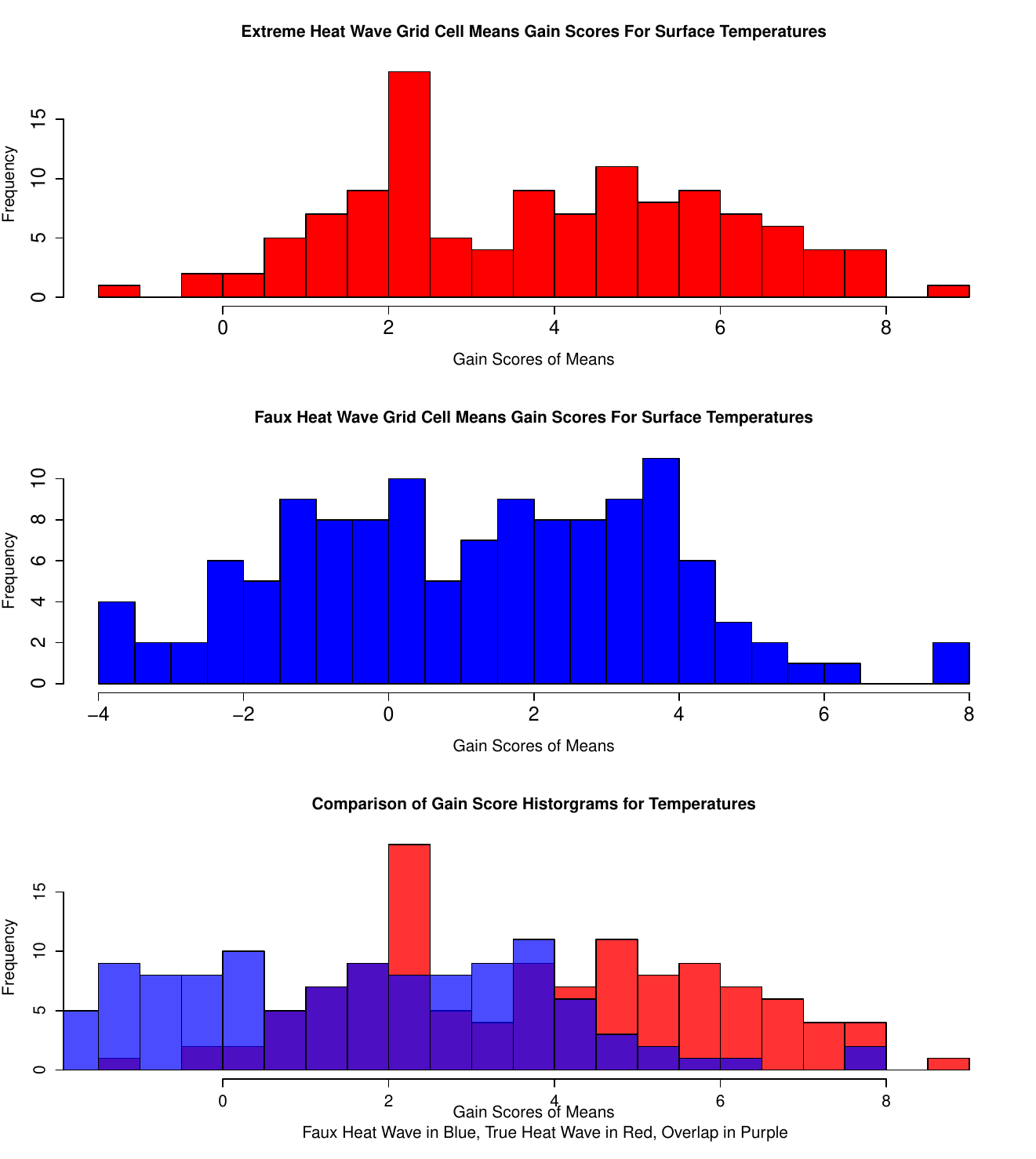}
\caption{fig: PNW Histograms of the gain scores of means for the extreme heat wave in red, faux heat wave 1 ind blue, and their overlap in purple.}
\label{fig:histos}
\end{center}
\end{figure}

Research on extreme heat waves often starts with comparisons between surface temperature distributions and common theoretical densities such as the extreme value distribution (Touchette, 2009; McKinnon and Simpson, 2022; Zhang et al., 2024)  
Figure~\ref{fig:histos} shows the grid cell, surface temperature gain scores of means histogram for the extreme heat wave, for the Faux Heat Wave 1 and for their overlap. The gain scores condition directly on pre-test surface temperatures and, therefore indirectly on fixed temperature-related features of grid cells such as proximity to an urban heat island.\footnote
{
Statistical interactions between dynamic processes and fixed features of a grid cell or the period (i.e., day) certainly can occur and are not controlled by the pre-test conditioning (Lui, 2016: section 1.4). Estimating such relationships can be very challenging and is beyond the goals of this analysis, necessarily constrained by what the AIRS data measures.
}
 
Neither the actual heat wave histogram nor faux heat wave histogram has much resemblance to the canonical normal distribution. There is far too much mass toward the centers of both histograms. Nor is either distribution strongly skewed to the left or the right. There is nothing to suggest an extreme value distribution. The attempt to isolate the role of dynamic temperature process does not lead to popular ``off-the shelf''  distributions in gain scores.\footnote
{
These are ``eyeball'' assessments that require some care because the aspect ratio of all three histograms produces rather flat-looking frequency profiles. When the histograms have an aspect ratio of 1 to 1 in height to width, the eyeball impression is close to a rectangle that is taller than it is wide. The 1 to 1 aspect ratio led to distorted displays when they were collected on a single page. 
}

One might conclude that the two histograms are much alike. However, the extreme heat wave histogram has 3 of 120 grid cells with negative values and an overall gain score mean of 3.8. The faux heat wave distribution has 44 of 128 grid cells with negative values and an overall gain score mean of 1.2.\footnote
{
There are less than 144 grid cells in either histogram because when a dataset for each was built with all of the predictors, there are 20 or more observations with missing data. Missing data can occur because of cloud cover or intermittent technical malfunctions. Missing observations in the AIRS data are an ongoing challenge.
}
One can explore further how the distributions compare by plotting the two histograms in the same axes. The bottom display shows the distributional offset in more detail. The red bars are taken from the histogram for the extreme heat wave gain scores. The blue bars are taken from the histogram for the faux heat wave 1 gain scores. The purple bars show the overlap. 

More is going on than a simple distributional shift to the right. The distributional shape changes as well. The extreme heat wave histogram has more of its mass above its mean flattened and extended toward higher gain score values. Nevertheless, there is a substantial overlap between the two distributions. Although extreme heat waves can be very consequential, a substantial fraction of the gain scores of means are much the same for both histograms. 

In the interest of space, Faux Heat Wave 2 is not yet discussed. But its histogram looks a lot like the histogram for Faux Heat Wave 1. We make more use of Faux Heat Wave 2 later to replicate some important findings for Faux Heat Wave 1. The key point now is that when examining histograms, the role of a 4 day lag versus a 30 day lead does not seem to make much empirical difference in the gain score histograms. 

\section{Some Important Predictor Distribution Differences Between the Actual Heat Wave and the Faux Heat Wave 1}\label{predictors}

One might wonder whether the actual heat wave and Faux Heat Wave 1 have dynamic precursors that differ as well.  We focus on dynamic predictors because only they can forecast a heat wave, which are also dynamic. Such precursors were listed in Table~\ref{tab:variables}. To proceed, however, we need first to consider a bit further how the rare events problem might be in part addressed.

\subsection{Mitigating The Rare Events Problem When Comparing Predictor Distributions}\label{mitigating}

The relative infrequency of extreme heat waves depends on the population from which the data were realized. For example, if one considers all days from 2017 through 2021 in the Pacific Northwest as the set of days from which the data were realized, there would be one realized extreme heat wave.

Labeling all four of the last days in June of 2021 as heat wave days, there would be 4 extreme heat wave days among more than 1800 days overall. Based on the data, if one never predicted an extreme heat wave, that prediction would be correct for over 99.8\% of the days without using any predictors at all.  It would be, therefore, almost impossible to uncover the roles of predictors because fitting is already close to perfect. Even if the days from which the data were realized included all just 2021, a forecast of no heat wave would be correct for 98.9\% of the days. The fitting problems are nearly as damaging. 

There is more promise considering only the 30 days in June of 2021. A no heat wave forecast will be right for ``only'' 86.7\% of the days. This degree of imbalance might be statistically manageable. We can make further progress employing a within-subject research design (Maxwell et al., 2018: section III) that uses the same grid cells as the study units for the post-test, pre-test, and precursors measured two weeks earlier. Figure~\ref{fig:histos} avoids in a transparent (even trivial) manner the rare event problem \emph{within} any of the three research design configurations because no comparisons between a true heat wave and a faux heat  wave are made.

There is also no evidence for rare event problem when making comparisons \emph{across} the three design configurations. This will become important later, especially in Section~\ref{sec:binary}. If the actual heat wave is compared empirically to either of the two faux heat waves, there are, in principle (i.e., without yet considering missing data), 144 true heat wave grid cells and 144 faux heat wave grid cells. If the extreme heat wave scenario is compared to a pooled version of the two faux heat waves, there remain 144 grid cells in the heat wave scenario and then, 288 grid cells in the combined faux heat wave scenarios. Because of the within-subject design, the PNW extreme heat wave in these data is not rare. 

There is a price to pay for these design remedies to the rare events problem. Perhaps most important, we use a form endogenous sampling when the three design configurations in Figure~\ref{fig:Designs_PNW} are imposed to all grid cells. It is unlikely that any study of rare extreme heat waves could be undertaken with data from a naturally-occurring population in which the probability that a realized case would be found in any one of three different study coditions is exactly $1/3$. Forcing the data to be balanced in this manner can produce misleading statistical results because of a mismatch between joint distribution of the data and the joint probability distribution of the population. (Manski and Lerman, 1977; Manski and McFadden, 1981). We will return to this problem later to consider several possible solutions. For now, we continue to descriptively unpack how extreme heat wave precursor distributions can differ from faux heat wave precursors distributions. Ideally, such information subsequently will help when real forecasting is undertaken.

\begin{figure}[htbp]
\begin{center}
\includegraphics[width=4in]{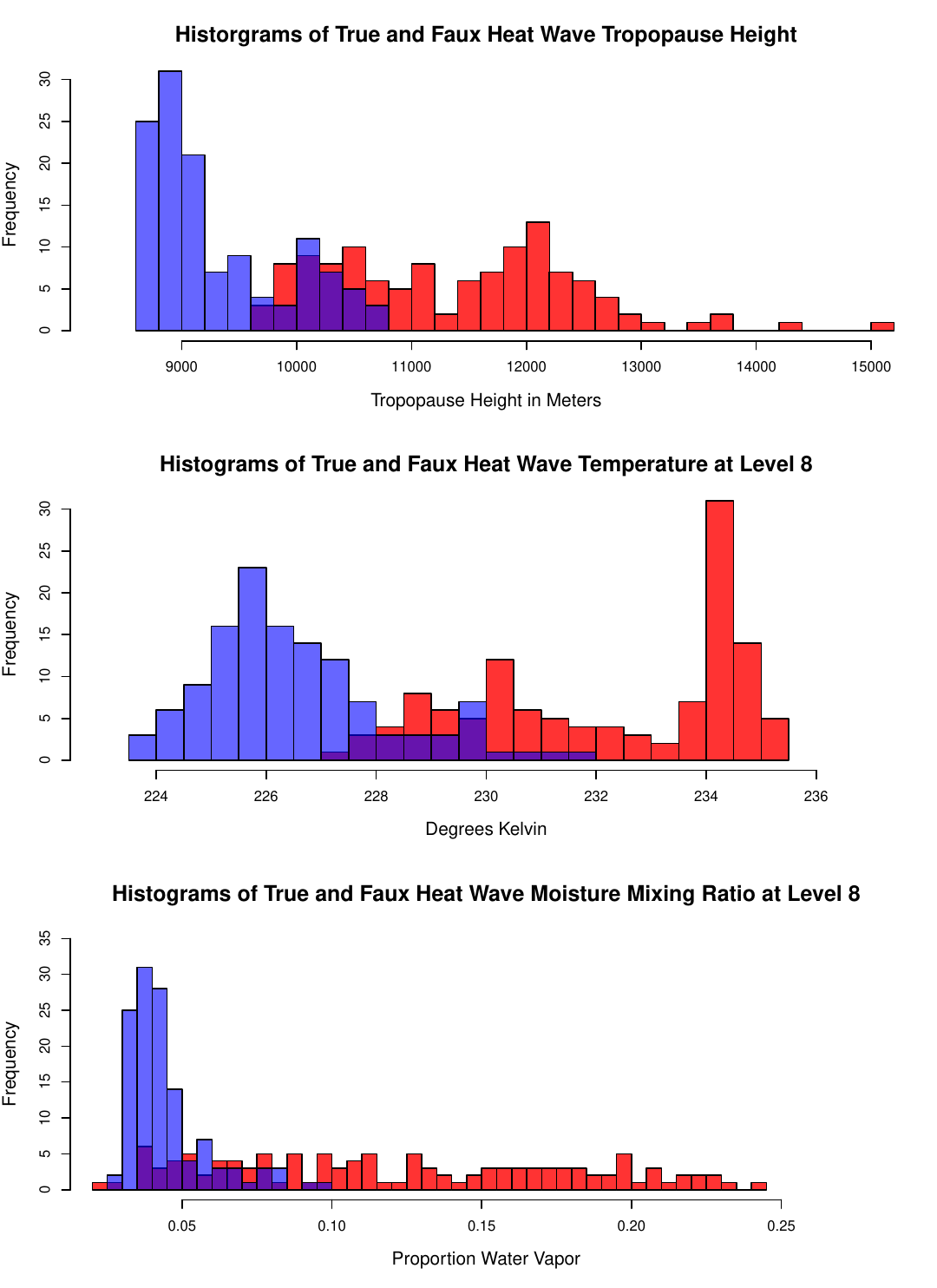}
\caption{Comparisons Between Three Important Predictors For The Actual Heat Wave and Faux Heat Wave 1}
\label{fig:predictors_PNW}
\end{center}
\end{figure}

\subsection{Some Predictor Comparisons Between the Actual Heat Wave and the Faux Heat Wave 1 or Faux Heat Wave 2}\label{sup}

Figure~\ref{fig:predictors_PNW} compares predictor histograms from the actual heat wave and Faux Heat Wave 1. At the top, is a comparison between distributions for tropopause height plotted red for the true heat wave, in blue for the Faux Heat Wave 1, and in purple for the overlap. The distributions differ in location, variability and shape. 14 days before both the actual heat wave or Faux heat wave 1, tropopause height was typically substantially higher in the actual heat wave grid cells. In the middle, there are also differences in distributional location, variability and shape. 14 days before both the actual heat wave or Faux heat wave 1, atmospheric temperature at level 8 was typically substantially higher in the actual heat wave grid cells. At the bottom, there are again distributional differences in location, variability and shape. 14 days before both the actual heat wave or Faux Heat wave 1, atmospheric water vapor at level 8 was typically substantially higher in the actual heat wave grid cells. 

We repeated the exercise but replaced Faux Heat Wave 1 with Faux Heat Wave 2. The faux heat wave now is 30 days after the true heat wave. Consequently, the risks from carry-over contamination are dramatically reduced. On the average, however, weather conditions in June and July may differ. There can be less rain and lower soil moisture in July, for instance. According to NOAA National Centers of Environmental Information, 1.9 inches of rain fell in Seattle in June of 2021, and 0.0 inches of rain fell in Seattle in July of 2021 (https://www.ngdc.noaa.gov/).\footnote
{
In the winter months, the rainfall in Seattle is much more substantial.
}

Despite such possibilities, the same comparisons produced similar results.
Perhaps neither the difference in pre-test to post-test separation time nor the differences in weather patterns between June and July matter much. It also may not matter if the actual heat wave followed a faux heat wave or the reverse. No evidence of carry-over, order problems is found.

There probably are many other variables having distribution differences depending on the circumstances in which summer temperatures materialize. Nevertheless, the comparisons between histograms offer some additional signs that extreme summer heat waves might differ structurally from ordinary summer heat. Work on quasiresonant amplification is a provocative illustration (Petoukov et al., 2012; Mann et al., 2018; Li et al., 2024). 

However, the empirical details are hard to pin down because there is substantial dependence between many such predictors, especially the when same variable is measured at different altitudes. Proximate values can be highly correlated. It is also important to not uncritically accept any answer to the ``compared to what" question. There are no doubt different kinds of ``non-heat-waves'' to which comparisons can be made. We picked comparisons by temporal and spatial proximity. There are perhaps better, or at least complementary, ways to proceed.

\section{Applying Genetic Algorithms To The Gain Scores}\label{GA}

The statistics literature has long suggested, inconclusively, ways to think about and analyze variables that are highly correlated (Alin, 2010). Some recent work seems especially relevant for our setting: variable selection combined with machine learning (Chan et al., 2022). But all approaches we have seen to date exploit the data on hand as the venue for progress. The recommendations are a long way from closure. 

If there were an \emph{ideal} extreme heat wave and an \emph{ideal} set of ordinary hot summer days, perhaps the empirical messiness of heat wave drivers could be reduced substantially. Recall that the June PNW faux heat wave had many gain scores similar to those of the June actual heat wave. This is not surprising because surface temperatures are produced by normal weather and topographical factors likely combined with dynamic forcings when a heat dome arrives. Might abstractions from extreme heat waves and usual summer days help? 

We borrow from the social sciences the concept of an ``idea type'' or a ``pure type'' (Garth and Mills, 1958) that abstracts the essential features of some phenomenon much like many key concepts in climate science. Then, perhaps a genetic algorithm can be employed to construct two ideal types: a synthetic population of extreme heat wave grid cells and a synthetic population of ordinary summer day grid cells. Comparisons between the two might allow for a more definitive reading of their similarities and differences.

 \subsection{Constructing Algorithmic Survival Functions}
 
 Genetic algorithms generally require as input a survival function.\footnote
 {
 ``Survival functions'' are also known as ``fitness functions'' or ``selection functions.''
 }
To construct a survival function separately for the heat wave gain scores and for the gain scores from Faux Heat Wave 1, we considered two forms of widely used and successful supervised learning: random forests (Breiman, 2001a) and stochastic gradient boosting (Friedman, 2001). Both can be seen as variants of nonparametric regression, are very easy to train, and often the default tuning parameter settings work well. A complication is that there is strong dependence between many of the AIRS predictors. For example, the measured temperatures at different altitudes are often correlated at values above .95. 

Random forests handles strongly correlated predictors relatively well because it is comprised of a large ensemble of regression or classification trees with the partitioning for each tree undertaken using a small random \emph{subsets} of predictors. Sampling predictors (without replacement) for each split can dramatically reduce the impact of high multicollinearity.\footnote
{
Random forests usually randomly samples cases when each tree is grown and predictors at each split in a tree. Random sampling of cases can be done with or without replacement. Random sampling predictors and cases is designed to increase the independence between trees. Greater independence facilitates the averaging over all trees in the forest. We recently learned that \emph{gbm3} in R, which implements various kids of boosting, now allows sampling predictors too, and along with its R predecessor \emph{gbm} allows random sampling of cases as well. In R, the choice between random forests (i.e., \emph{randomForests}) gradient boosting now depends largely on details that are peripheral to this discussion.
}
We settled on the implementation of random forests in R, ported by Andy Law and Matthew Wiener (Liaw and Weiner, 2002) from the original Fortran by Breiman and Cutler (Breiman, 2001a), which includes several important auxiliary algorithms that can dramatically increase algorithmic transparency.\footnote
{
We briefly considered using deep neural networks through the keras3 package in R authored and maintained by Tomasz Kalinowski. Unlike random forests algorithm, neither the predictors nor the observations are randomly sampled. The vastly different units for some of the predictors caused problems for the back propagation. Dependencies between the predictors also caused computational difficulties. Once both problems were addressed, tuning was able to proceed. For all practical purposes, the fitting performance was about as good as random forests'. But, some of the auxiliary procedures available for random forests, which shine some light into the algorithmic black box (Molnar, 2022), were not readily available. For example, the widely applicable partial dependence plots procedure \textit{pdp} in R, authored and maintained by Brandon Greenwell, apparently requires that a custom prediction function be coded when used on neural networks output.
}

A random forest was trained using 500 regression trees for the June extreme heat wave. The same gain score data were used for training as in the earlier gain score comparisons in Section~\ref{histo}. The trained and averaged ensemble of regression trees serves as a survival function for a genetic algorithm's construction of an ideal, synthetic population of grid cells experiencing a true heat wave. 

In the same manner, a random forest was trained as using 500 regression trees on June faux heat wave 1. The trained and averaged ensemble of regression trees serves as a different survival function for a genetic algorithm's construction of an ideal, synthetic population of grid cells experiencing a faux heat wave.

\subsection{An Application of a Genetic Algorithm To Construct An Idea Type
True Heat Wave and and Ideal Type Faux Heat Wave} 

The two survival functions were used as input to the procedure GA in R written by Luca Scrucca (2013). The standard defaults for numeric predictors worked well. For both survival functions, 5000 iterations were specified leading to 5000 synthetic populations of 100 grid cells each having increasing fitness on the average over iterations. As a first approximation, one can think of a genetic algorithm as jittering predictor values, computing new fitted values from the survival function using the jittered predictors, saving the subset of the grid cells with the  larger gain scores, and repeating this process many times until there is a population that includes only very fit grid cells (i.e., having large gain scores). In the process, predictor dependence can be reduced substantially because of the randomness the genetic algorithm introduces to predictors.\footnote
{
The GA software in R has several tuning parameters with good default values.(1) For each synthetic population of grid cells, the predictors from the five grid cells with the highest computed surface temperatures gain scores were not altered. (2) The cross-over probability for each possible pair of predictors was set at 0.8. A cross-over refers to a pair of ``parent'' grid cells combining their predictor values to form an ``offspring'' grid cell.  (3) Because all of the predictors were treated as numeric in this application, the predictor values for an offspring grid cell were computed as the average of the corresponding predictors from a random pair of parent grid cells. (4) The mutation probability for each grid cell's predictor was set at 0.1. (5) The mutation method used replaced the value of the existing numeric variable with a value within a specified range drawn at random from a uniform distribution. The size of a synthetic population is not a tuning parameter but 100 grid cells seemed a good compromise between an ability to examine the output visually in detail and having a sufficient number of observations to arrive at sufficiently stable parameter estimates.
} 

\subsubsection{Genetic Algorithm Results}

The grid cell mean gain score for the final synthetic population produced by the actual heat wave survival function was 7.26 degrees K. The grid cell mean gain score for the final synthetic population produced by the faux heat wave 1 survival function was 5.76 degrees K. Because of differences in the two survival functions, the genetic algorithm did not equate the two. This is another vehicle through which to reflect on how the physics for a true heat wave may differ fundamentally from the physics of a faux heat wave.

\begin{table}[htp]
\footnotesize
\centering
\caption{Genetic algorithm predictor ``solution'' values for true heat wave
and the faux heat wave 1.}
\vspace{5pt}
\begin{tabular}{|c|c|c|}
\hline \hline
Predictor & True Heat Wave & Faux Heat Wave 1 \\
\hline
metersabove  &   623.02  &     395.78  \\   
land       &     0.73  &    0.67    \\
temp4    &       274.48    &   228.16 \\     
temp5    &       268.84    &   232.32 \\     
temp6    &       259.13    &    224.15 \\     
temp7     &      246.98     &   225.02 \\     
temp8     &      233.93     &   211.24 \\     
temp9     &      215.74     &   245.85 \\     
temp10   &       217.13  &   260.26 \\    
temp11   &       218.14     &  227.19 \\     
 temp12   &      220.84     &  265.94 \\    
mix4        &   5.33  &       6.97 \\     
mix5        &  3.44  &       1.47 \\      
mix6        &  1.67  &       0.76 \\    
mix7        &   0.61 &      0.16 \\    
mix8        &   2.60 &     6.86 \\     
mix9        &   1.31 &     5.84 \\     
mix10      &   4.47 &    3.90 \\  
mix11      &  7.68  &    5.73 \\     
mix12      &  2.84  &   5.20 \\     
tropheight & 11766.76 & 6728.78 \\
\hline \hline   
\end{tabular}
\label{tab:solution}
\end{table}

Table~\ref{tab:solution} provides more evidence about the ways in which an ideal true heat wave can differ from an ideal faux heat wave. Shown are the predictor values the algorithm fashioned for the final population that produced grid cells with superb fitness in the metric of surface temperature gain scores. The predictor values for the extreme heat wave are in the left column. The predictor values for Faux Heat Wave 1 are in the right column.  Comparisons between the two columns can be used for variable selection in a simple application of feature engineering.\footnote
{
This is an example of cherry picking that can invalidate conventional statistical inference. One is engaged, even if indirectly, in model selection (Berk et al., 2013). However, analyses to follow are able sidestep this problem.
}

Compared to the ideal faux heat wave, the ideal true heat wave is characterized by (1) substantially higher grid cell elevation (i.e. 623 meters versus 396 meters), (2) a higher tropopause (i.e. 11,767 m versus 6,729 m), (3) higher grid cell atmospheric temperatures at lower altitudes, and (4) lower grid cell atmospheric temperatures at higher altitudes. The role of the atmospheric water vapor mass mixing ratios is inconsistent. There is now perhaps a bit better sense of which predictors are related most strongly  to an extreme heat waves compared to a faux heat waves. 

A minor complication is that grid level elevation and location over land or water are both fixed. They cannot be directly related to the extreme heat wave. The genetic algorithm was applied a second time, but the two fixed variables were not included. There were no material changes in the conclusions compared to those from Table~\ref{tab:solution}, and the mean gain scores for the final populations were effectively the same as well. 

\section{Extreme Heat Wave Classification Using a Binary Outcome}\label{sec:binary}

With the long term goal accurate forecasting of extreme heat waves two weeks (or more) in advance, several practical hurdles exist. The predictors must be measured and readily available well in advance of a possible heat wave. This implies data collection on a regular basis. There are also several second order requirements. For example, the data collection requires continuity in naming conventions, variables measured, the temporal and spatial scales, the measurement technology used, and data file formats. If any of these change, forecasting skill can be badly damaged. At the very least, abrupt nuisance variation in the data over time will likely be introduced.

In addition, a pre-test is no longer feasible. Any pre-test immediately before a possible heat wave makes forecasting unworkable. When a forecast is needed two weeks in advance, the pre-test values would not yet have materialized. Data curation and associated research designs must adjust.

 Moreover, a numeric outcome such as surface temperature should be revisited.
 An extreme heat wave is a discrete event; it is either present or not present. Therefore, the task becomes forecasting a binary outcome. When using earlier temperature gain scores, we were fitting a \emph{consequence} extreme heat wave precursors and perhaps the impact of a heat dome. We were not explicitly forecasting a discrete, extreme heat wave. Regression should be replaced by classification.
 
 Finally, as an intermediate step in the direction of real forecasting, it should be possible for predictors positioned substantially earlier to identify accurately real heat waves compared to faux heat waves.  But accurate classification does not guarantee accurate forecasting; for classification the outcome is known whereas when forecasting, the outcome is not known. (If it were known, there would be no need for a forecast.) New statistical complications can arise.
 
 \subsection{Research Design Revisions}

Returning to the design in Figure~\ref{fig:Designs_PNW} for the June true extreme heat wave, we drop the pre-test. We also recode the post-test to equal ``1.''  For purposes of comparison, we use July, 2021 as the setting for a faux heat wave (i.e, Faux Heat Wave 2). The pre-test is dropped and the post-test is recoded to equal ``0.'' The same 144 grid cells are used for a new extreme heat wave design configuration and for the new faux heat wave design configuration; the same grid cells are exposed to one climate regime in June and then another climate regime in July. The June data are then ``stacked'' above (or below) the July data to form a single dataset. The classification format is balanced, and the rare event problem seems resolved. 

But the rationale goes a bit deeper. When the data are stacked, each study unit no longer serves as its own control. What remains, however, is that the \emph{composition} of the study units does not change. The same set of grid cells are exposed to precursors of a true heat wave and to precursors of the faux hear waves. The grid cells serve as the their own \emph{collective} controls. The confounding would be far more challenging if one set of grid cells were exposed to true heat wave precursors and another set of grid cells were exposed to faux heat wave precursors. 

We use the last 4 days of July for the faux heat wave because the time offset between the actual and faux heat wave is 30 days. The chance of carry-over effects becomes very small. This helps justify the absence of a pre-test. Because the earlier results for Faux Heat Wave 1 and Faux Heat Wave 2 were so similar, the different seasonal conditions between June and July probably do not matter much for our classification analysis. 

\subsection{Classification Using Random Forests}

The earlier comparisons between predictor distributions for a subsequent true heat wave or a subsequent faux heat wave, coupled with the genetic algorithm output, promise excellent classification results. We considered two forms of widely used and successful supervised learning classifiers: random forests (Breiman, 2001a) and stochastic gradient boosting (Friedman, 2001). Both can be seen as variants of nonparametric regression, are very easy to train, and often the default tuning parameter settings work well. A complication is that there is strong dependence between many of the AIRS predictors. For example, the measured temperatures at different altitudes are often correlated at values above .95. 

Random forests handles strongly correlated predictors relatively well because it is comprised of a large ensemble of regression or classification trees with the partitioning for each tree undertaken using a small random \emph{subsets} of predictors. Sampling predictors (without replacement) for each split can dramatically reduce the impact of high multicollinearity.\footnote
{
Random forests usually randomly samples cases when each tree is grown and predictors at each split in a tree. Random sampling of cases can be done with or without replacement. Random sampling predictors and cases is designed to increase the independence between trees. Greater independence facilitates the averaging over all trees in the forest. We recently learned that \emph{gbm3} in R now allows sampling predictors too, and along its R predecessor \emph{gbm} also allows random sampling of cases. In R, the choice between random forests (i.e., \emph{randomForests}) gradient boosting now depends largely on details that are peripheral to this discussion.
}
We settled on the implementation of random forests in R, ported by Andy Law and Matthew Wiener (Liaw and Weiner, 2002) from the original Fortran by Breiman and Cutler (Breiman, 2001a), which includes several important auxiliary algorithms that can dramatically increase algorithmic transparency.\footnote
{
We briefly considered using deep neural networks through the keras3 package in R authored and maintained by Tomasz Kalinowski. Unlike random forests algorithm, neither the predictors nor the observations are randomly sampled. The vastly different units for some of the predictors caused problems for the back propagation. Dependencies between the predictors also caused computational difficulties. Once both problems were addressed, tuning was able to proceed. For all practical purposes, the fitting performance was about as good as random forests'. But, some of the auxiliary procedures available for random forests, which shine some light into the algorithmic black box (Molnar, 2022), were not readily available. For example, the widely applicable partial dependence plots procedure \textit{pdp}, authored and maintained by Brandon Greenwell, apparently requires that a custom prediction function be coded when used on neural networks output.
}

Random forests was trained as a classifier using the new data structure. All of the predictors used previously were candidates for the classification exercise. As a start, however, we selected three predictors from Table~\ref{tab:solution} that were dynamic and seemed most promising: (1) height of the tropopause, (2) atmospheric water vapor mass mixing ratio at level 8, and (3) the temperature at level 8. For each, we considered the gaps shown in Table~\ref{tab:solution} between values in the synthetic heat wave column and values in the synthetic faux heat wave column. For the mixture and temperature variables, other proximate altitudes could have been selected because proximate altitudes tend to be substantially correlated. Dependence between the three selected predictors by themselves was not a terribly serious complication but still risks some unstable results. We emphasize again that an algorithm is not a model and even if it were, that model would be badly misspecified. There is also the matter of cherry picking the predictors.
 
 The random forests results are rather strong. Overall classification error rate is approximately 5.0\%. In this setting, a false positive can be defined as incorrectly classifying the faux heat wave as the true heat wave. The false positive rate is 5.7\%. In this setting, a false negative can be defined as incorrectly classifying the true heat wave as the faux heat wave. The false negative rate is about 4.1\%. In short, the random forest classifier almost always correctly identifies the true positives and the true negatives (i.e., very high sensitivity and specificity respectively).  

\begin{figure}[htbp]
\begin{center}
\includegraphics[width=2.7in]{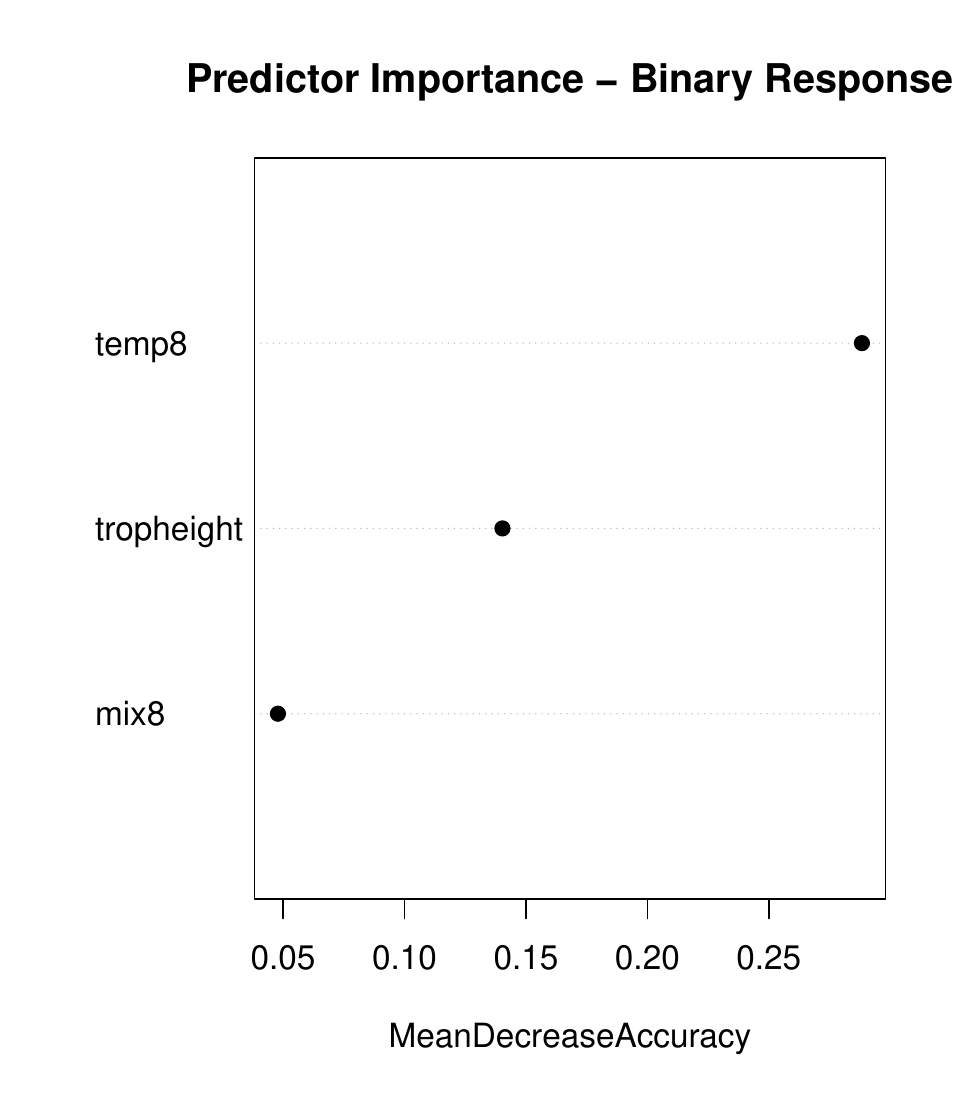}
\caption{Variable importance for accurate classification of heat waves. The variable names were shortened for ease of writing code but are more fully explained in the text.}
\label{fig:ForImp}
\end{center}
\end{figure}

Figure~\ref{fig:ForImp} is the variable importance plot from random forests. The metric at the bottom of the figure is the reduction in classification accuracy averaged over both outcome classes when each variable in turn is randomly shuffled with the other variables fixed at their existing values. For example, classification accuracy declines by nearly 30 percentage points if temperature at altitude 8 (only) is shuffled. In short, Figure~\ref{fig:ForImp} shows that all three predictors are important for classification accuracy.\footnote
{
Average classification accuracy over both outcome classes is used because the numbers of false negatives and false positive are very similar. There is at this point no need to distinguish between the two outcome classes.
}

\begin{figure}[htbp]
\begin{center}
\includegraphics[width=4.5in]{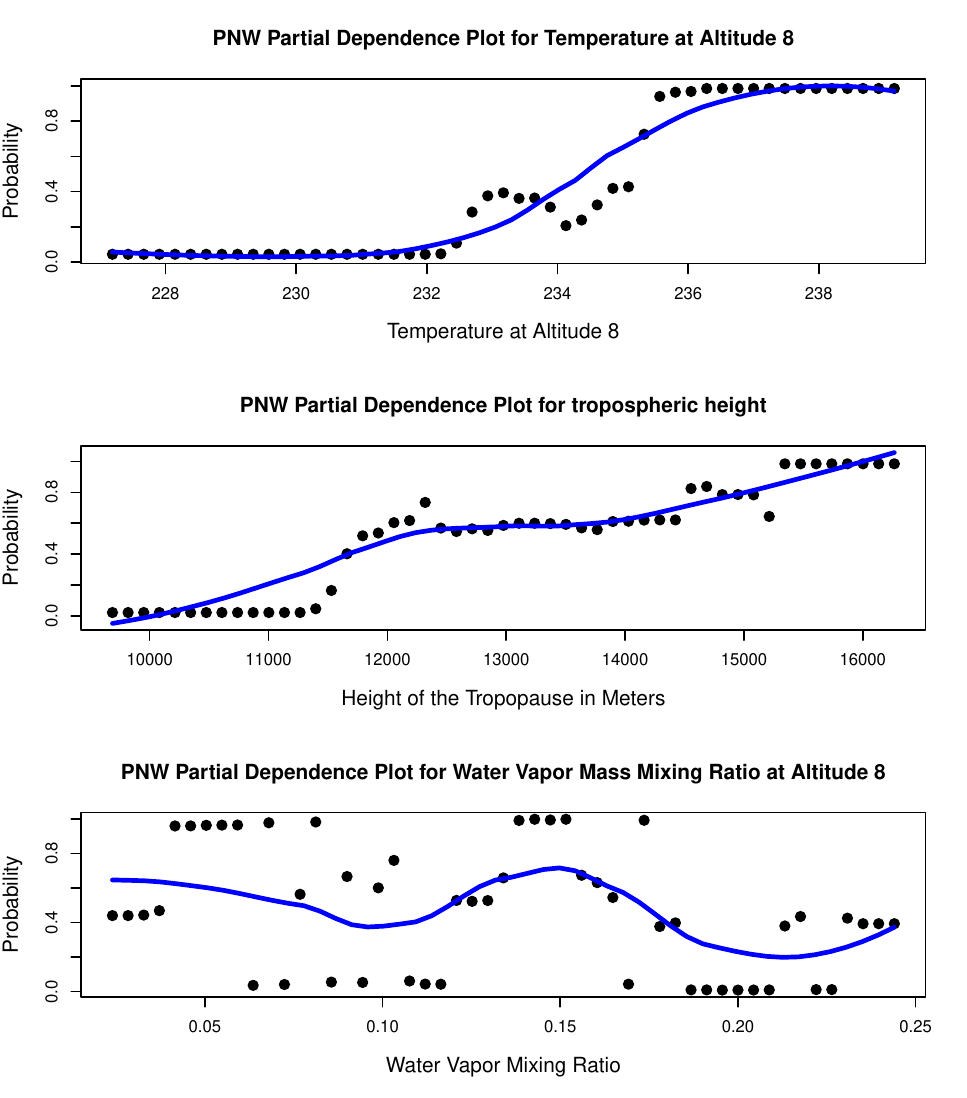}
\caption{Partial Dependence Plots for the PNW Forecasts.}
\label{fig:pdpforecasts}
\end{center}
\end{figure}

Figure~\ref{fig:pdpforecasts} show the partial dependence plots for the three predictors. On each vertical axis is the fitted conditional probability of the extreme heat wave for different values of a given predictor, all other predictors fixed at their mean values in the data (Friedman, 2001; Greenwell, 2017).\footnote
{
The probabilities range from 0.0 to 1.0 with fitted values needing much of that range. Placing all three plots in the same figure (in the interests of space) attenuates visually the strength of the associations summarized by the smoother.
} 
The possible impact of correlated predictors used by the classifier is controlled, but not with conventional covariance adjustments used parametric classification (e.g., logistic regression). The blue line overlaid is a lowess smoother (Cleveland and Devlin, 1988).

The partial dependence plot at the top is for the predictor atmospheric temperature at altitude 8. Judging by the smoothed fitted values, there is an S-shaped relationship with the probability of a heat wave, but the fitted values are actually flat until the temperature is approximately 232 K and flat again after the temperature reaches about 235 K. Sparse data in the tails of the temperature distribution may be the explanation. Overall, the relationship is positive. Fitted probabilities range from near 0.0 to near 1.0 with temperature increases from two weeks earlier. 

The partial dependence plot in the middle is the height of the tropopause in meters. Judging by the smoothed fitted values, there is as the height increases nearly a straight line increase in the probability of a heat wave from about a 0.0 probability to almost a 1.0 probability. However, fitted values are flat until atmospheric pressure reaches 11500 meters, perhaps an impact of sparse data for lower atmospheric pressures.

The bottom partial dependence plot shows the relationship between heat wave probabilities and the atmospheric water vapor mass mixing ratio at altitude 8. Consistent with its weaker importance to the fit, the relationship shown by the smoother seems unsystematic. The fitted values appear disorderly. However, the disappointing performance does not seem to undermine the overall classification accuracy when all three variables are used at once. 

In summary, the classification exercise produced superb accuracy. Clearly, the predictors work even with a 2 week lag. The variable importance plot and the partial dependence plots help to make sense of the accuracy. However, our extensive exploratory examination of the data places a thumb on accuracy scale. The empirical results are valid on their face, but in other settings, absent the such a rich preliminary examination of the data, classification accuracy could be substantialy lower.

\subsection{Forecasting Uncertainty}\label{uncertainty} 

As a statistical matter, high classification accuracy can support high forecasting accuracy. However, in addition to the complications just described, we picked the true heat wave dates before the classification analysis began. Our results are conditional on first knowing the dates of the true heat wave and the faux heatwaves and then assigning the appropriate outcome classes homogeneously to all grid cells.  When the grid cells had been exposed two weeks earlier to extreme heat wave precursors, each grid cells was coded as ``1.''  When the grid cells had been exposed two weeks earlier to faux heat wave precursors, each grid cells was coded as ``0.'' That makes forecasting even more problematic. In particular, one cannot properly take the next step into valid forecasting. 

One should ordinarily expect proper forecasts coupled with estimates of uncertainty. A good option for discrete outcomes is conformal prediction sets that produce legitimate forecasts and valid uncertainty estimates even in finite samples. The key assumption is exchangeability (Gupta et al., 2022). Under the popular split samples approach, the data are randomly separated into two disjoint samples: (1) a training dataset that a classifier can use for fitting, and (2) a calibration dataset used construct prediction sets. A wide variety of classifiers can be used be they algorithms or models. There is no requirement that either contains all of the relevant predictors or employs the appropriate functional forms. Still, for unlabeled cases needing a forecast, a conformal prediction set will include a forecast at a predetermined coverage probability properly interpreted as the probability that the forecast is correct. No assumptions need be made about the joint probability distribution of the dataset or additional properties of the calibration data. 

But, one must be able to make a credible case that at least the calibration data are exchangeable. This is difficult requirement for the PNW data. Because we selected the outcomes in advance, the realized data are not IID or exchangeable, and the calibration data would not be exchangeable either. In addition, the algorithm specification incorporated by the classifier is the product of various kinds of prior analyses raising post-model-selection inferential challenges (Berk et al., 2013; Kuchibhotla et al., 2022). 

\section{A Replication Study in Phoenix, Arizona}\label{replication}

Repeating the clasification exercise in a new geographical setting can have several benefits. First, one is able to consider whether the statistical approaches employed earlier generalize well. Does the formulation of a particular data generating process more broadly apply? Are gain scores instructive elsewhere? Are the random forests and the \emph{GA} genetic algorithms appropriate in other locations? Second, there is substantive generalizability. In particular, using the same variables with their values realized elsewhere, how well reproduced is the confusion table from the Pacific Northwest? All of these questions can be properly addressed with no feature engineering, no ``bake-offs'' between algorithms, no model selection, no p-hacking, and no filtering of reported results. 

\subsection{Data Challenges}

We obtained AIRS daily data for the Phoenix area for 63 grid cells from 2019 through 2023. Data were not available for 2024. The measured variables were the very same used in the PNW analysis. However, the climate and topography in the greater Phoenix area are very different from the PNW and far more homogeneous. The Pacific Northwest is lush. The greater Phoenix area largely is a desert. We picked the greater Phoenix area because of the strong replication challenges it posed. 

We had about half the number of grid cells for the Phoenix area than for the Pacific Northwest. There were also a substantial number of grid cells with missing data. Between the two, we anticipated a much smaller number of observations overall. One result could much weaker statistical power coupled with still more sparsity. We sought a procedure to increase the number of usable grid cells for the within-subject (i.e., grid cells) approach.

According to many media accounts, the Phoenix heat dome arrived at the beginning of July 2023, not at end of June, 2021, as the extreme heat wave  had in the Pacific Northwest. We proceeded much as before, but faux heat waves A and B, representing ordinary summer temperatures, were positioned in early July of 2021 and early July of 2022 respectively, well before the actual Phoenix extreme heat wave. The predictors lagged 14 days earlier fell toward the middle of June in 2021, 2022 and 2023. For all three Phoenix years, we used a binary response variable coded “1” for the extreme heat wave in early July of 2023, and “0” for faux extreme heat waves in early July of 2021 and early July 2022. The actual heat wave apparently began the around first day July, 2023. Correspondingly, the two faux heat waves began on the first day of July 2021 and the first day of July 2022.

All three time periods were then stacked to make single data set. Each grid cell in principle was observed in three different years and in principle, there would be 189 observations overall, each exposed to one set of extreme heat wave precursors and two sets of faux heat wave precursors. One benefit is that the collection of study subjects was the same under all three conditions; there could be no changes in sample composition. 

\begin{figure}[htbp]
\begin{center}
\includegraphics[width=4in]{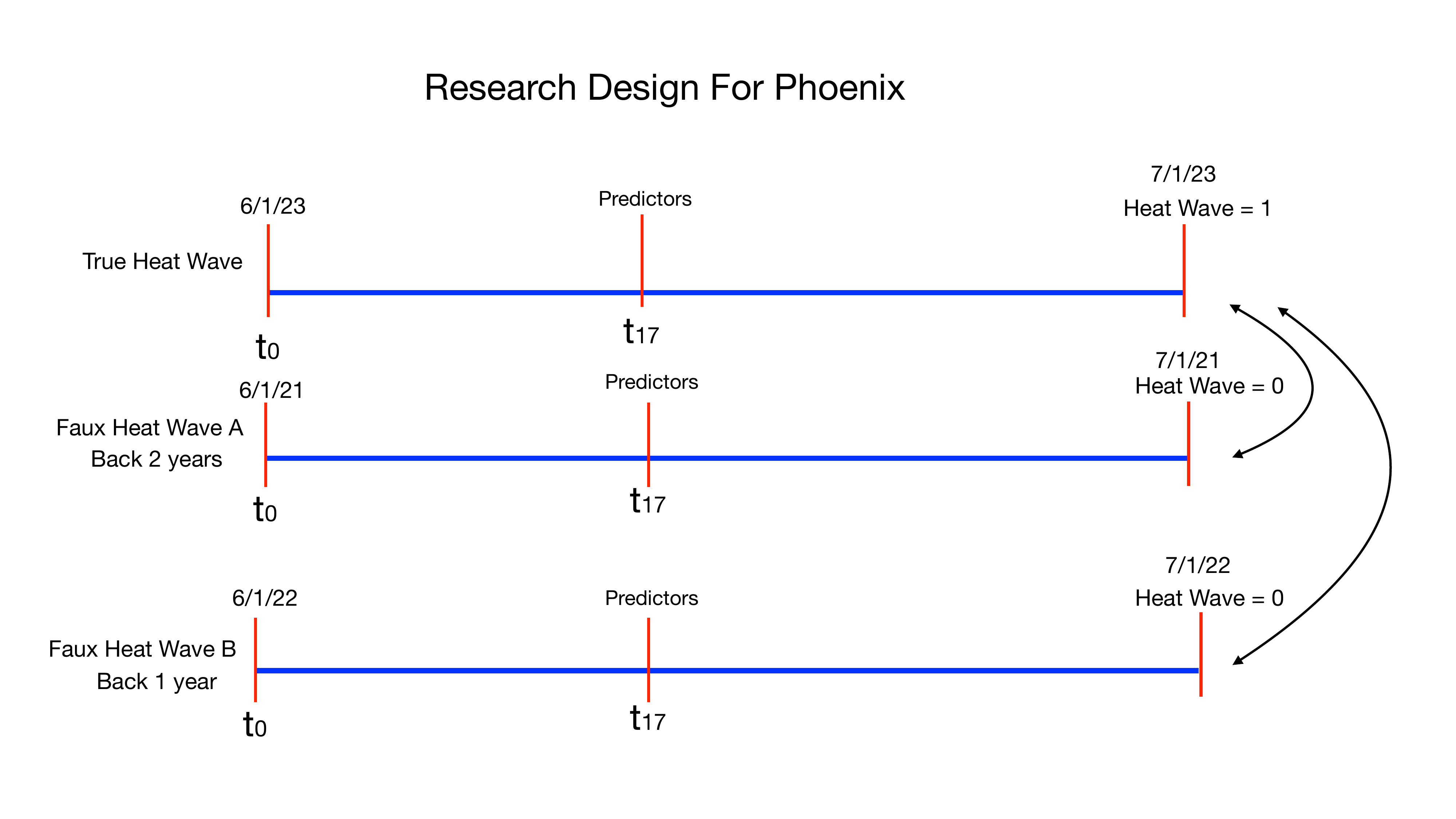}
\caption{Research Design Configurations for the Greater Phoenix Forecasts}
\label{fig:designphoenix}
\end{center}
\end{figure}

But the potential problems from missing data remained. When we included only the three precursor variables required for PNW comparability, the number of grid cells with complete data fell to 128. Because the data were lost through cloud cover and occasional downtime on the remote sensing satellite, one might argue that the data were missing in blocks, conditionally at random. Nevertheless, the large amount of missing data raises concerns about the representativeness of the Phoenix data and about substantial instabilities that could be introduced. There would be further complications should there be differential amounts of non-random missingness when the grid cells were exposed to true heat wave precursors versus the when the grid cells were exposed to the faux heat wave precursors. The design structure is shown in Figure~\ref{fig:designphoenix}.\footnote
{
Because the data were likely generated in an unknown, spatially dependent matter, all of the missing data imputation methods with which were familiar seem inappropriate.
}

\begin{figure}[htbp]
\begin{center}
\includegraphics[width=4.5in]{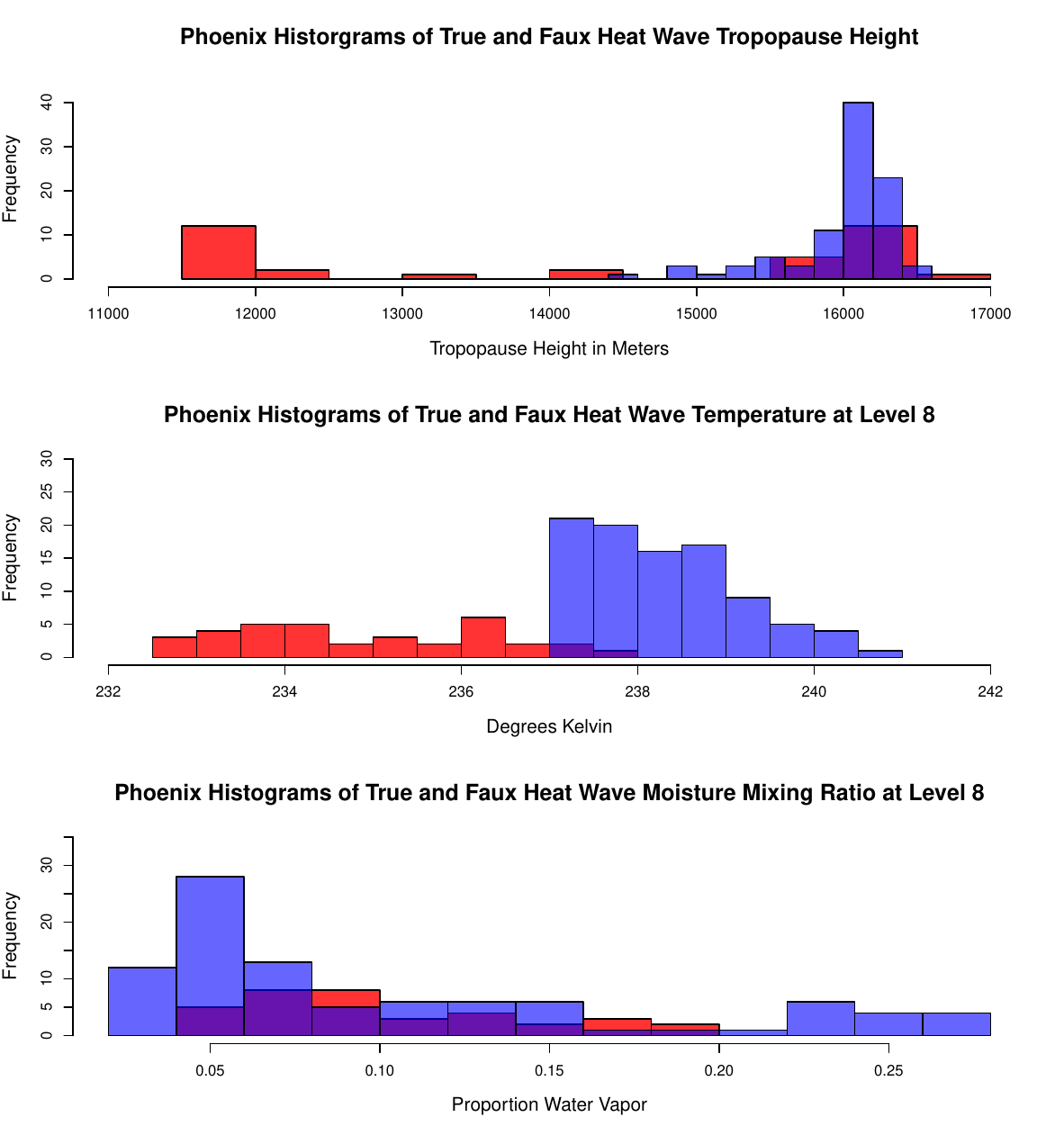}
\caption{fig: Phoenix Histograms for Predictors, Actual Heat Wave in red, faux heat waves ind blue, and their overlap in purple.}
\label{fig:histphoenix}
\end{center}
\end{figure}

\subsection{Comparisons Between True Heat Wave and Faux Heat wave Predictor Histograms}

As before, we compare the distributions for each of three predictors when exposed to the true heat wave precursors compared to the faux heat wave precursors. The amount of missing observations are more pronounced for the true heat wave data. We expected about 33\% of the data to be for the true heat wave precursors, and about 67\% of the data to be for the faux heat wave precursors. After listwise deletion, the proportions were 27\% and 73\% respectively. The corresponding numbers of observations were 35 and 93. True heat wave cases are noticeably underrepresented, and the sample size is small enough to be worrisome.

For the true heat wave and faux heat waves, Figure~\ref{fig:histphoenix} shows in the top two plots different predictor distributions for tropopause height and for temperature at level 8. But in contrast to the PNW, the true heat wave predictor values tend to fall toward the \emph{left} of the faux heat wave values. The bottom plot for the water vapor mixing ratio, is again inconclusive. Compared to the PNW data, perhaps because of the larger fraction of missing values from the true heat wave precursors, greater tropopause heights and higher temperatures are disproportionately lost from the red bars. That might account for some results that at this point seem counterintuitive.

\subsection{Classification Results}

We began simply. Using the random forests classifier already trained with data from the Pacific Northwest, we obtained the fitted probabilities with input data from the greater Phoenix area. That is, the trained algorithm was fixed. The data from Phoenix were employed solely to construct fitted probabilities. Whatever the relationships captured for the PNW were being implicitly assumed to apply in Phoenix. 

The resulting Phoenix confusion table has an overall misclassification proportion of a little over .16, which is about 3 times larger the overall misclassification proportion for the PNW. There are no false positives, but the false negative fraction is 40\%. This substantial difference in the proportion of false negatives compared to the proportion of false positives may be in part a result of the marginal distribution of the response variable. Whereas there are for the PNW data nearly the same number of actual heat wave precursor exposures as faux heat wave precursor exposures, the proportions for Phoenix have approximately two faux heat wave precuresor exposures for each actual heat wave precursor exposure. With weighting methods described in the next section, the empirical ratio of false negatives to false positives could be made to approximate 1 to 1, but the complications that can follow would need to be addressed.

A second comparability assessment was undertaken using the random forest classifier just as before, but with the Phoenix data structured as in Figure~\ref{fig:designphoenix}.  All three predictors used in the PNW analysis are used again. After deletion of all rows with any missing data, there were 128 grid cell observations. With so few observations, substantial instability was expected. The implicit assumption was that the same precursors would matter, but probably not in the same way as before. 

We were pleasantly surprised by the confusion table produced. The overall misclassification rate for Phoenix is a little over .03, about the same as error rate for the PNW. There were no false positives, and a false negative rate of approximately .11  The difference was likely explained in part by the relative decline in the number of in the heat wave cases, now accounting for just a little over a quarter of the data. All three predictors contribute substantially to the fit, although the role of atmospheric temperature 14 days earlier is more dominant than for the PNW.\footnote
{
Given the likely instability because of the modest number of training data observations, we examined the confusion tables following from 10 additional random splits of the data. Overall misclassification proportions ranged from a little over .05 to a little under .12. Often there were no false positives. By most standards, classification accuracy was quite good but lacked the stability desired.
}

\begin{figure}[htbp]
\begin{center}
\includegraphics[width=4in]{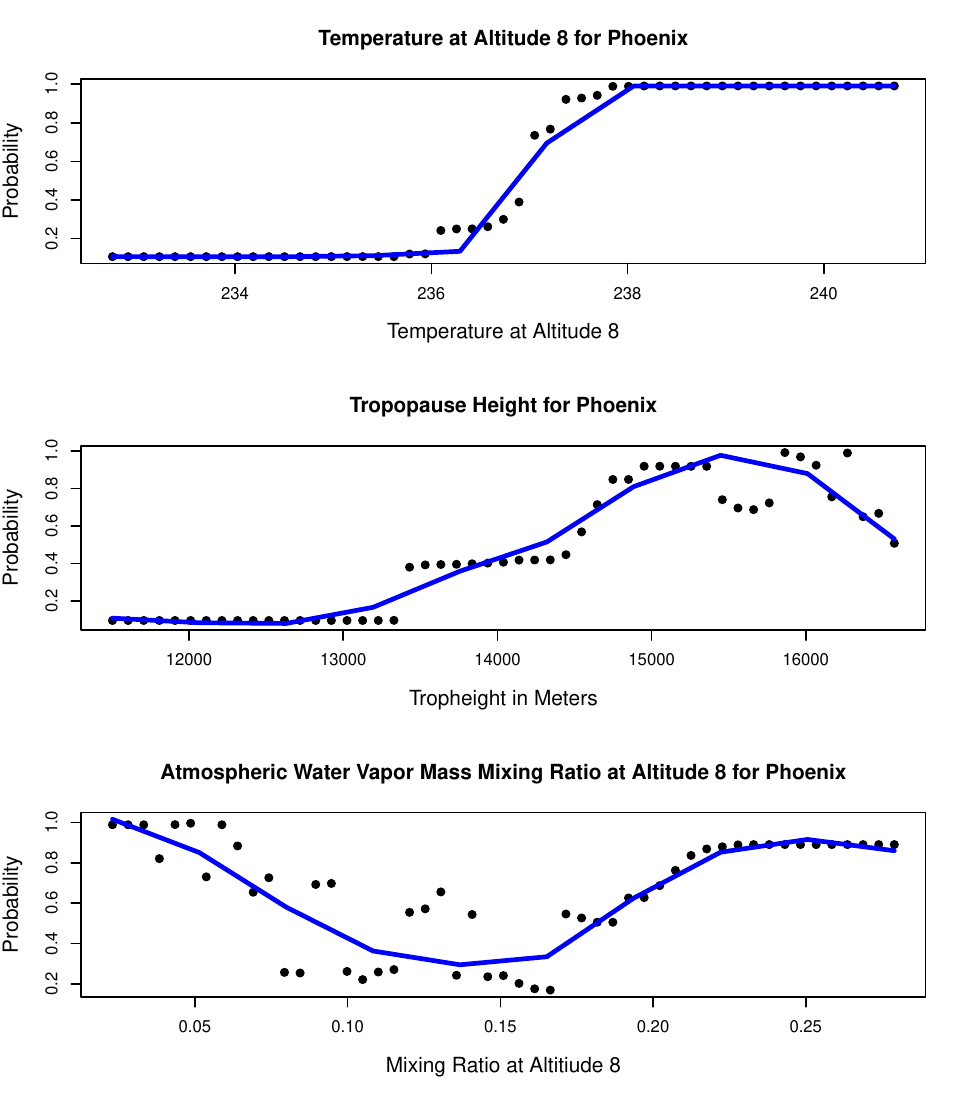}
\caption{Partial Dependence Plots for Phoenix Forecasts}
\label{fig:pdpphoenix}
\end{center}
\end{figure}

Despite the histogram differences just described and great differences in climate and topography, the partial dependence plots in Figure~\ref{fig:pdpforecasts} for the PNW and for Figure~\ref{fig:pdpphoenix} for the greater Phoenix area have smoothers showing much the same S-shaped relationship between the fitted probabilities and atmospheric temperature at an altitude of 8.\footnote
{
As before, the vertical variation in the plots is visually attenuated.
}
Both relationships begin flat for their lowest values before their slopes turns positive, and end flat at the highest values. The Phoenix relationship is shifted a bit to the right in response to its higher temperatures. It also somewhat less smooth because of data that are more sparse.

Comparing the PNW and Phoenix smooths for tropopause height, both start out flat before turning positive. For the PNW,  the overall pattern is roughly linear and positive thereafter. For the greater Phoenix area, the relationship smoother turns negative at around 15,500 meters, and the tropopause heights start as somewhat higher values.

For the water vapor mixing ratio at altitude 8, neither the smoother for the PNW nor for the greater Phoenix area show easily interpreted relationships. This helps explain why in both geographical settings, the water vapor mix by itself contributes relatively little to the fit. The good news is that it seems not to be needed. 

The apparent contradiction between the predictor comparative histograms and the partial dependence plots is perhaps explained by the difference between marginal and conditional distributions. The partial dependence plots are conditioned so that adjustments are made for the dependence between the three predictors. For example, in the Phoenix data, the correlation between temperature at level 8 and tropopause height is .83. 

\begin{figure}[htbp]
\begin{center}
\includegraphics[width=3in]{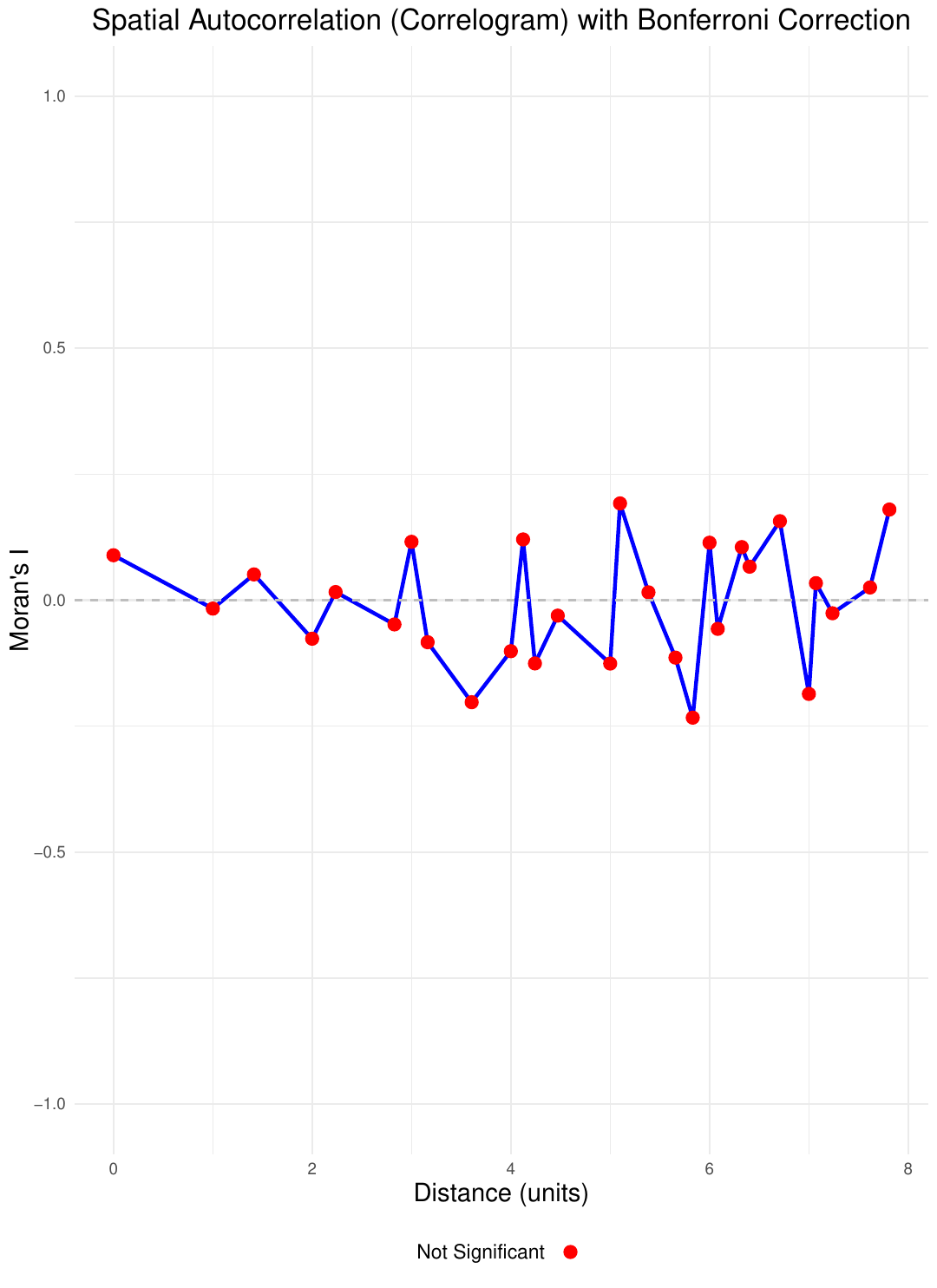}
\caption{Spatial Correlogram at Different Distances in Units of .01 with Statistical Tests for Each Subject to the Bonferroni Correction for the Nominal Critical Value of .05.}
\label{fig:spatial}
\end{center}
\end{figure}

Despite the excellent classification accuracy, one might wonder if there remain dependencies in the Phoenix analysis. Given the spatial nature of the data, there could remain spatial dependence between grid cells implying relationships not yet addressed. Classifier residuals were computed. The spatial correlogram R procedure \emph{gstat}, authored and maintained by Edzer Pebesma, was applied. Figure~\ref{fig:spatial} shows the result. The correlogram shows that the measure of association, here Moran's I, fluctuates around 0.0 and is not systematically related to the distance between spatial residuals.\footnote
{
The residuals can in principle range between $\pm 1.0$. The discrete unit for the plot is .01. The horizontal axis is in distance multiples of .01.
}
After a Bonferroni correction for multiple testing, for none of the associations is the null hypothesis of no association rejected. The nominal critical values was .05.

It is important to emphasize that although the PNW analysis was steeped in extensive data exploration, the Phoenix analysis had none. Yet classification accuracy is similar for both, and two of the three partial dependence plots are roughly similar in across the two locations. But as before, conformal prediction sets cannot properly be used. Again, we knew the days on which the Phoenix heat wave occurred prior to the data analysis, and grid cells in advance were coded homogeneously. The data are neither IID nor exchangeable.

\section{Problems from Endogenous Sampling}\label{endogenous}

The within-grid-cell  designs for the PNW and for the greater Phoenix area led to excellent balance that remedied the rare events problem for our binary outcome of heat wave or not. However, it is challenging to think of a naturally occurring population of grid cells that could be a plausible source of the data in which such balance would have occurred. Our curation process substantially oversampled extreme heat wave days, which is a form of \emph{endogenous} oversampling. 

As noted earlier, there is a literature on problems that can result when data collection alters the prior distribution of the response variable compared to its distribution in the population of interest (Manski and Lerman, 1977; Manski and McFadden, 1981). For example, in a study of cancer detection in lung X-rays, one might oversample cancer cases compared to cancer-free cases. Absent the oversampling, there would be too few ``positives'' for a sufficiently accurate detection assessment (Wang et al., 2021). But the price is badly distorted estimates of the sensitivity and specificity for the cancer diagnoses.

For numeric response variables, clever sampling strategies have been proposed that, in effect, oversample for extreme values (Ragone et al., 2018). Endogenous sampling is explicit and worrisome. For discrete response variables, our focus here, several related remedies have been proposed (Cosslett, 1993; Waldman, 2000). A dataset can be weighted in different ways so that the imbalance is corrected, and the endogenous (response) variable once again can provide a valid estimates. Sometimes the weighting is buried in the statistical machinery along with other performance enhancements, but it is still altering the sample imbalance. For example, when Matthew's correlation, also know as the phi coefficient, is used as an objective function for a classifier to maximize, it can help correct imbalance (Dourard it al., 2024). But because the cost ratio target of false positives to false negative is set at 1.0, the marginal distribution is, in effect, being re-weighted.

A  simple form of appropriate weighting is easily undertaken. For example, whether or not there is an extreme heat wave can be the endogenous (response) variable as shown in Figure~\ref{fig:designphoenix}. Suppose $P(heat)$ is the probability of an extreme heat wave in the target population, and  $P^\ast(heat)$ is the probability of an extreme heat wave in the curated data on hand.  The two probabilities are different because of the manner in which the data were collected and organized. There is a pair of remedial weights (Manski and Lerman, 1977; Manski and McFadden, 1981). For each grid cell $i$ that experienced an extreme heat wave, $weight_{heat,i} = \frac{P(heat_i=1)}{P^\ast (heat _i = 1)}$. For each grid cell $i$ that did not experience an extreme heat wave, $weight_{noheat,i} = \frac{P(heat_i=0)}{P^\ast (heat _i = 0)}$. These weights can be incorporated easily within a weighting option commonly available in machine learning software.\footnote
{
As before, a heat wave is 	coded equal to ``1.'' The absence of a heat wave is coded equal to ``0.''
} 

There are several difficulties with such solutions. To begin, in many applications it is not apparent what the proper population is or even should be. An earlier example was whether the population for studying a rare extreme heat wave should be daily ARIS data for the most recent 5 years, daily AIRS data for the most recent year, or daily AIRS data for the most recent month. 

In addition, for rare discrete events, the weighting proposals typically recommended can restore the very problem that the initial oversampling of rare events tried to remedy. Their compensatory down-weighting makes the formerly rare events rare once again. Preferred machine learning software may not run at all, or may run with unsettling warning messages. For example, the fitted probabilities may fall outside of the 0 to 1 range. 

Some adjustments back toward the target population can trigger deeper difficulties. Therneau and Atkinson (2023) show that for tree-based classifiers such as classification trees, the prior distribution of the response is related to the ratio of the number of false negatives to the number of false positives (or the reciprocal). That ratio is often used to represent the relative costs of false negatives to false positive (i.e., all costs, not just monetized costs). For example, if for a binary response variable a classifier produces 40 false positives and 20 false negatives, there are 2 false positives for every false negative. This can be interpreted to mean that \emph{false negatives} are twice as costly as false positives, which may help explain why the classifier fits the data so that there are relatively fewer of them; one false negative is ``worth'' two false positives. The costs are asymmetric.

In response to asymmetric costs, there are weighting adjustments leading to what Kuhn and Johnson (2013: Chapter 6) call ``cost sensitive training.'' Related thinking can be found in focal loss approaches (Lin et al., 2017) and implicitly in gradient boosting (Friedman, 2001). Rednick and his colleagues (2021) offer the most involved approach that weights more heavily the hard to classify observations, and a specified outcome class can be given a higher priority as well.\footnote
{
We coded their loss function in keras3 in R. It ran, but did far worse that the usual  cross-entropy loss, perhaps because the classifier was already too successful with the within-subject design and the balanced data that resulted. There were too few false negatives and false positives usefully to weight. We also found tuning a conceptual challenge when we tried to return to a highly imbalanced source population response variable combined with a plausible stakeholder consensus cost ratio.
}  
In their classic book, Luce and Raiffa (1957) formulate the problem as risk minimization rather than loss minimization. Broadly framed, asymmetric loss functions are required to account for asymmetric fitting costs.
  
The conflation of corrections for sample imbalance with the estimated misclassification cost ratio is easily illustrated. Using the expressions above for the weights, the PNW classification exercise was repeated. Recall that the response variable was binary, and there were the same number of grid cells that experienced true heat wave precursors as experienced faux heat wave precursors. Using the month of June, 2021 as the target population, for a little more than 13\% of the days, the 144 grid cells experienced the extreme heat wave precursors. For the other days, the 144 grid cells the precursors were for a faux  heat wave.  In the data analyzed, for 50\% of the days, the 144 grid cells experienced extreme heat wave precursors. A disparity between 13\% and 50\% may be more extreme than anticipated by Manski and Lerman (1977), but we can proceed for illustrative purposes.

After weighting, the population imbalance was approximately reproduced. Before weighting, the classification error rate was a little over 5\%, and there were nearly the same number of false negatives as false positives. The cost ratio was about 1.0, conveying equal costs. When Manski-Lerman weighing was applied, the classification error rate rose to nearly 14\%, and the cost ratio of false positives to false negatives increased from about 1.0 to nearly 10.0. False positives were now almost 10 times more costly than false negatives. In other words, the costs of heat wave false alarms were substantially up-weighted compared to the costs of true alarms not sounded. Many stakeholders might strongly prefer the reverse. In short, weighting changes the prior distribution of the response variable \emph{and} the relative costs of false negatives to false positives.

For the classification results reported for the PNW and the greater Phoenix area, we had no access to stakeholders nor any concrete policy context. The numbers of false negatives and false positives were very few and of roughly equal size. Classifier training determined an empirical cost ratio of about 1 to 1. Going forward in a real policy setting combined with stakeholder input, the within-subject research design could be revised to better capture practical concerns. The main requirement is that grid cells continue to capitalize on the within-subject structure. Additional faux heat waves could easily be added so that a more appropriate, but manageable, imbalance was achieved. Another possibility might be to train the classifier on balanced data but to improve fitted values, use as input an altered dataset with many more faux heat waves. 

\section{Summary and Conclusions}

Our efforts to provide a foundation from which to forecast extreme heat waves were guided by several key decisions. To begin, within the supervised learning traditions used for forecasting, the main response variable used in training was a binary label indicating whether an extreme heat wave had occurred or not. With so many extant definitions of heat waves, we settled on an operationalization emphasizing a heat dome and responses to extreme heat waves that are often dichotomous: preparations to open public buildings with air conditioning as ``cooling centers," having first responders ready to check on elderly people often living alone, making preparations to close public schools in the afternoons giving ample time for parents to arrange for child care, requiring for water breaks at outdoor work sites, re-allocating staff in emergency rooms to handle increased patient volume, and stocking emergency rooms and urgent care centers with adequate medical supplies (e.g., IV fluids, electrolyte solutions, ice packs, heat-related medication), and much more. 

We relied on the AIRS remote sensing data. These data do not depend on climate simulations of any kind including downscaled global climate models. This seems especially important because of our focus on rare, extreme events. AIRS data are probably noisier than simulation data and provide fewer variables with which to work; the data have promise but offer special challenges. Despite the challenges, the data in the PNW and in the Phoenix area appear to deliver instructive results.

Although our work has been informed by the climate science literature, we make no explicit use of models. Our analyses rest on quasi-experimental designs, a supervised learning algorithm well suited for forecasting, and a genetic algorithm for simple feature engineering. Accurate forecasting is our primary long term goal. Any explanatory results are a secondary benefit. 

Consistent with our extensive exploration of the data, the classification results in the Pacific Northwest and in the greater Phoenix area proved to be extremely accurate. Although the variable importance plots and partial dependence plots were not identical in the two locations, a case can be made that they are qualitatively similar despite important differences in local climate, topography, research designs and data quality. Still, many more study sites surely are needed, ideally with many more grid cells per site.

By taking genuine forecasting seriously, forecasting uncertainty had to be taken seriously as well. We capitalized thinking for nested conformal prediction sets that can provide guarantees of valid coverage even in finite samples as long as the calibration data are exchangeable. However, the manner in which the data were curated violated exchangeability. Future work can put true forecasting in play by not conditioning on the particular days in which extreme heat waves occurred. 

We acknowledged the issue of endogenous sampling, but argued that corrections for response variable imbalance are only part of the problem. The cost ratio of false negative and false positive classification errors is implicated as well. Sample imbalance and misclassification cost ratios are mathematically linked and could imply the need for an asymmetric loss function. We briefly considered some other options but currently lack the requisite information from a real policy setting having engaged stakeholders. To start the discussion, our results are based on within-subject designs that achieve adequate sample balance and a cost ratio of about 1 to 1. 

Even if our work proves helpful, it is at best a start. Among the more important issues going forward is our choice of a two week lag between forecasting precursors and heat wave outcomes. We relied on recent research, but a two week lag is hardly definitive. Our extreme heat wave operationalization also needs to be revisited. Empirically, it seems to work well, but others might work well too.There are many options already in print. Another significant matter is formulating appropriate populations that are instructive and yet provide better balanced data. We suspect that even if promising populations are proposed, endogenous sampling complications will be ongoing. Extreme heat waves may be increasing, but they are still relatively rare, and  ``compared to what'' questions will remain. Also remaining are research designs and analysis that undertake valid forecasting. Finally, there are many steps between hand-tailored research procedures and production quality forecasting tools that can be routinely used to deliver accurate heat dome warnings sufficiently in advance. We did not address such matters here. That would be at least premature.

\section*{References}
\begin{description}
\item
Alin, A. (2010)  ``Multicollinearity.'' \textit{Wiley Interdisciplinary Reviews: Computational Statistics} 2(3): 370 -- 374.
\item
 Aumann, H.H., Chahine, M.T.,Gautier, C.,Goldberg, M.D., Kalnay, E.,McMillin, L.M., Revercomb, H., Rosenkranz, P.W., Smith, W.L., Staelin, D.H., Strow, L.L., Susskind, J. (2003) ``AIRS/AMSU/HSB on the Aqua Mission: Design, Science Objectives, Data Products, and Processing Systems.'' \textit{Transactions on Geoscience and Remote Sensing} 41(2): 253 -- 264.
\item
Aumann, H.H., Broberg, S., Manning, E.M.,  Pagano, T., and Sutin, B. (2020) \textit{AIRS Level 1C Algorithm Theoretical Basis} Jet Propulsion Laboratory.
\item
Barriopedro, D., Garcia-Herrera, R., Ordonez, C., Miralles,  D.G., and Salcedo-Sanz, S. (2023) ``Heat Waves: Physical Understanding and Scientific Challenges.'' \textit{Reviews of Geophysics} 61: 1 -- 54.
\item
Bartusek, S., Kornhuber, K., and Ting, M. (2022) ``2021 North American Heatwave Amplified by Climate Change-Driven Nonlinear Interactions.'' \textit{Nature 
Climate Change} 12: 1143 -- 1150.
\item
Benjamin-Chung, J., Arnold, B.F., Berger, D., Luby, S.P., Miguel, E., and Colford Jr., J.M. (2018) ``Spillover Effects in Epidemiology: Parameters, Study Designs and Methodological Considerations.'' \textit{International Journal of Epiudemiology} 47(1): 332 - 347.
\item
Berk, R.A., (2018) \textit{Machine Learning Risk Assessments in Criminal justice Settings.} Springer.
\item
Berk, R.A., Brown, L., Buja, A., Zhang, K. and Zhao, L. (2013) ``Valid Post-Selection Inference.'' \textit{The Annals of Statistics} 42(2): 802 -- 837.
\item
Berk, R.A., Buja, A., Brown, L., George, E., Kuchibhotla, A.K., Su, W. and Zhao, L. (2019) ``Assumption Lean Regression.'' \textit{The American Statistician} 75(1); 76 -- 84.
\item
Box, G.E.P., Jenkins, G.M.,  Reinsel, G.G. and Ljung, G.M. (2016) \textit{Time Series Analysis: Forecasting and Control.,} Third Edition. Wiley.
\item
Breiman, L. (2001a) ``Random Forests.'' \textit{Machine Learning} 45: 5 --32.
\item
Breiman, L. (2001b) ``Statistical Modeling: Two Cultures.'' \textit{Statistical Science} 16(3): 199 -- 231.
\item
Burger, G., Sobie, S.R., Cannon, A.J., Werner, A.T., and Murdock, T.Q. (2013) ``Downscaling Extremes: An Intercomparison of Multiple Methods for Future Climate.'' \textit{Journal of Climate} 26(10): 3429 -- 3449. 
\item
Campbell, D.T, and Stanley, J.C. (1963) \textit{Experimental and Quasi-Experimental Designs for Research}. Houghton Mifflin.
\item
Chan, J.Y-L., Leow,  S.M.H., Bea,  K.T., Cheng, W,K., Phoong, S.W., Hong,  Z-W., and Chen, Y-L. (2022) ``Mitigating the Multicollinearity Problem and Its Machine Learning Approach: A Review.'' \textit{Mathematics} 10(8): 1283. https://doi.org/10.3390/math10081283.
\item
Chernozhukov, V., Wuthrich, K., and Zhu, Y. (2018) ``Exact and Robust Conformal Inference Methods for Predictive Machine Learning with Dependent Data.''  \textit{Proceeding of Machine Learning Research} 75: 1 -- 17.
\item
Chil\`es, J.-P., and Delfiner, P. (2012) \textit{Geostatistics: Modeling Spatial Uncertainty}, second edition. Wiley.
\item
Cleveland, W. S., and Devlin, S. J. (1988) "Locally Weighted Regression: An Approach to Regression Analysis by Local Fitting." \textit{Journal of the American Statistical Association} 83(403):  596 -- 610.
\item
Coles, S. (2001) \textit{An Introduction to Statistical Modeling of Extreme Values} Springer.
\item
Cosslett, S.R. (1993) ``Estimation from Endogenously Stratified Samples.'' \textit{Handbook of Statistics} 11: 1 -- 43. Elsevier. doi.org/10.1016/S0169-7161(05)80036-7
\item
Alessandro Dosio, A., Mentaschi, L., Fischer, E.M., and Wyser, K. (2028) ``Extreme Heat Waves Under 1.5 ◦C and 2 ◦C Global Warming.'' \textit{Environmental Research Letters} 13: 054066.
\item
Douard, N., LaRovere, J., Parmar, V., Straulino, D., Harris, M.D., McCann. C., Nasseri.A., Hanson C., Bataglia, M., Afshin E.E., Filiaci,  M., Azzarelli, K.,  Giakos, G., and Ebby Elahi, E. (2024) ``Multimodal Machine Learning for Multifactor Geospatial Prioritization of Suitable Healthcare Facility Sites.''   \textit{IEEE} IST International Cpnference on Imaging Systems \& Techniques. 
\item 
The Economist (2024) ``The Rise of a Truly Cruel Summer.''  June 26, 2024.
\item
Fischer, E.M., Beyerle, U., Bloin-Wibe, L., Gessner, C., Humphrey, V., Lehner, F. A., Pendergrass, A.G., Sippel, S., Zeder, J., and  Knutti, R. (2023) ``Storylines for Unprecedented Heatwaves Based on Ensemble Boosting.'' \textit{Nature Communications} 14(4643): 112 -- 114.
\item
Fontana, M., Zeni, G., and Vantini, S. (2023) ``Conformal Prediction: A Unified Review of Theory and New Challenges.'' \textit{Bernoulli} 29(1): 1 -- 23.
\item
Freedman, D.A. (2009) \textit{Statistical Models: Theory and Practice} Cambridge.
\item
Friedman, J.H. (2001) ``Greedy Function Approximation: A Gradient Boosting Machine",
 "The Annals of Statistics.'' \textit{The Annals of Statistics} 29(5): 1189 -- 1232.
 \item
 Fuller, W.A. (1987) \textit{Measurement Error Models} Wiley.
 \item
Gebrechorkos, S., Leyland, J., Slater, L., and coauthors (2023) ``A High-Resolution Daily Global Dataset of Statistically Downscaled CMIP6 Models for Climate Impact Analyses.'' \textit{Scientific Data} 10: 611. https:// doi.org/10.1038/s41597-023-02528-x
 \item
Gettleman, A., and Rood, R.B. (2016)
 \textit{Demystifying Climate Models: A Users Guide to Earth System Models} Springer.
 \item
Gilleland, E., and Katz, R.W. (2016) ``extRemes 2.0: An Extreme Value Analysis Package in R.'' \textit{Journal of Statistical Software"} 72(8) 10.18637/ jss.v072.i08.
\item
Gupta, C., Kuchibhotla, A.K., and Ramdas, A. (2022) ``Nested Conformal Prediction and Quantile Out-of-bag Ensemble Methods.'' \textit{Pattern Recognition} 127: 108498. https://doi.org/10.1016/j.patcog.2021.108496.
\item
Greenwell, B.M. (2017) ``pdp: An R package for Constructing Partial Dependence Plots.'' \textit{The R Journal} 9(1): 421 -- 436. 
\item
Hastie, T.J., and Tibshirani, R.J. (1990) \textit{Generalized Additive Models}. Chapman and Hall.
\item
Jain, P., Sharma, A.R., Acuna, D.C., Abatzoglou, J.T. and  Flannigan, M. ``Record-breaking Fire Weather in North America in 2021 Was Initiated by the Pacific Northwest Heat Dome. Communications Earth \& Environment.'' Published Online 5(202). https://doi.org/10.1038/s43247-024-01346-2.
\item
Jones, B.,  and Kenward, M.G. (2014) \textit{Design and Analysis of Cross-Over Trials}. CRC Press.
\item
Karniadakis, G.E., Kevrekidis, I.G., Lu, L., Perdikaris, P., Wang, S., and Yang, L. (2021) ``Physics-Informed Machine Learning.'' Nature Review Physics 3: 422 -- 440. https://doi.org/10.1038/s42254-021-00314-5
\item
Kearns, M., and Roth, A. (2019) \textit{The Ethical Algorithm: The Science of Socially Aware Algorithm Design.} Oxford University Press.
\item
Khatana, S.A.M., Szeto, J.J., Eberly, L.Aa, Nathan, A.S., Puvvula, J., and Chen, A. (2024) ``Projections of Extreme Temperature–Related Deaths in the US.'' \textit{JAMA Network Open} 7(9): e2434942.  doi:10.1001 jamanetworkopen.2024.34942.
\item
Kotz, S., Balakrishnan, N., and Johnson, N.L. (2000) \textit{Continuous Multivariate Distributions,} Volume1. Wiley.
\item
Kuchibhotla, A.K., Kolassa, J.E., and Kuffner, T.A. (2022) ``Post-Selection Inference.'' \textit{Annual Review of Statistics and Its Applications} 9: 505 --527.
\item
Kuchibhotla, A.K. and Berk, R.A. (2023) ``Nested Conformal Prediction Sets for Classification with Applications to Probation Data.'' \textit{Annals of Applied Statistics} 17(1): 761 -- 785.
\item
Kuhn, M., and Johnson, K. (2013) \textit{Applied Predictive Modeling}. Springer.
\item
Lafferty, D.C. and Sriver, R.L. (2023) ``Downscaling and Bias-Correction Contribute Considerable Uncertainty to Local Climate Projections in CMIP6.'' \textit{npj Climate and Atmospheric Science.} 6(158) https:// doi.org/10.1038/s41612-023-00486-0.
 \item
 Li, Y., Wu, X., Yang, P., Jaing, G., and Luo, Y. (2022) ``Machine Learning for Lung Cancer Diagnosis, Treatment, and Prognosis.'' \textit{Genomics Proteomics Bioinformatics} 20(5): 850 -- 866.
 \item
Li, X. , Mann, M.E., Wehnerb, M.F., Rahmstorf, S., Petric, S., Christiansena, S. and Carrilloa, J." (2024) ``Role of Atmospheric Resonance and Land–Atmosphere Feedbacks as a Precursor to the June 2021 Pacific Northwest Heat Dome event.'' \textit{PNAS: Earth Atmospheric, and Planetary Sciences} 121(4): 1 -- 7.
\item
Liaw, A. and Wiener, M. (2002) ``Classification and Regression by randomForest.'' \textit{R News} 2/3: 18 -- 22.
\item
Lin, T-y., Goyal, P., Girshick, R.B., He, K., and Doll\'{a}r, P. (2017) ``Focal Loss for Dense Object Detection.'' arXiv:1708.02002 [cs.CV]
\item
Luce, R.D., and Raiffa, H. (1957) \textit{Games and Decisions: Introduction \& Critical Survey} Wiley
\item
Lui, K-J. (2016) \textit{cross-over Designs: Testing, Estimation, and Sample Size.} Wiley.
\item
Ma, Y., Chen, S., Ermon, S., and Lobell, D.B. (2024) ``Transfer Learning in Environmental Remote Sensing.'' \textit{Remote Sensing of Environment} 301: 113924.
\item
Mann, M.E., Rahmstorf, S., Kornhuber, K., and Steinman, B.A. (2018) ``Projected Changes in Persistent Extreme Summer Weather Events: The Role of Quasi-Resonant Amplification.'' \textit{Science Advances} 4(10) DOI: 10.1126/sciadv.aat3272.
\item
Manski, C.F., and Lerman, S.R. (1977) ``The Estimation of Choice Probabilities from Choise Based Samples.'' \textit{Econometrica} 45(8) 1977 -- 1988.
\item
Manski, C. and McFadden, D. (1981). “Alternative Estimators and Sample Designs for Discrete Choice Analysis,” in C. F. Manski and D. McFadden (eds.) \textit{Structural Analysis of Discrete Data with Econometric Applications.} MIT Press.
\item
Maris, E. (1998) ``Covariance Adjustment Versus Gain Scores—Revisited." \textit{Psychological Methods} 3(3): 309 -- 327.
\item
Marx, W., Hunschild, R., and Bornmann, L. (2021) ``Heat Waves: A Hot Topic in Climate Change Research.'' \textit{Theoretical and Applied Climatology} 146: 781 -- 800.
\item
Maxwell, S.E., Delaney, H.D., and Kelly, K. (2018) \textit{Designing Experiments and Analyzing Data.} Routledge.
\item
McKinnon, K. and Simpson, I.R. (2022) ``How Unexpected Was the 2021 Pacific Northwest Heatwave?'' \textit{Geophysical Research Letters} 49(18) (e2022GL100380).
\item
Mearns, L.O., Bukovsky, M.S., Pryor, S.C., and Magana, V. (2014) ``Downscaling of Climate Information.'' In G. Ohring. \textit{Climate Change in North America. Regional Climate Studies.} Springer.
\item
Molnar, C. (2022) ``Interpretable Machine Learning: A Guide for Making Black Box Models Explainable.'' Independently published and available on Amazon.
\item
Oliveira, R.I., Orenstein, P, Ramos, T., and Romano, J.V. (2022) ``Split Conformal Prediction for Dependent Data.''   https://arxiv.org/abs/ 2203.15885.
\item
Perkins, S.E. (2015) ``A Review on the Scientific Understanding of Heatwaves—Their Measurement, Driving Mechanisms, and Changes at the Global Scale. \textit{Atmospheric Research} 164 -- 165: 242 -- 267.
 \item    
Petoukhov, V., Rahmstorf, S., Petri, S., and Schellnhuber, H.J. (2013) ``Quasiresonant Amplification of Planetary waves and Recent Northern Hemisphere Weather Extremes.'' \textit{Proceedings of the National Academy of Sciences} 110: 5336 -- 5341.
\item
Piticar, A., Cheval, S., and Frighenciu, M. (2019) ``A Review of Recent Studies on Heat Wave Definitions, Mechanisms, Changes, and Impact on Mortality.''\textit{Forum Geographic} XVIII(2): 103 -- 120.
\item
Ragone, F., Wouters, J., and Bouchet, F. (2018) ``Computation of Extreme Heat Waves in Climate Models Using a Large Deviation Algorithm.'' \textit{Proceedings of the National Academy of Sciences} 155(1): 24 -- 29.
\item
Ridnik, T., Ben-Baruch, E., Zamir, N., Noy, A., Friedman, I., Protter, M., and Zelnick-Manor, L. (2021) ``Asymmetric Loss for Multi-Label Classification.'' \textit{Proceedings of the IEEE/CVF International Conference on Computer Vision} (ICCV), 2021 :82 -- 91.
\item
Rinaldo, A., Wasserman, L., and G'Sell, M (2019) ``Bootstrapping and Sample Splitting For High-Dimensional, Assumption-Lean Inference.'' \textit{The Annals of Statistics} 47(6): 3438 -- 3460.
\item
Rockel, B. (2015) ``The Regional Downscaling Approach: a Brief History and Recent Advances.'' \textit{Current Climate Change Reports.} 1 22 -- 29.
 \textit{Proceedings of the 33rd International Conference on Neural Information Processing Systems} (NeurIPS), Vancouver, Canada.
\item
 Schmidt, G. (2024) ``Why the 2023's Heat Anomaly Is Worrying Scientists.'' \textit{Nature} 267(21): 467.
 \item
Schneider, T. and Behera, S. and Boccaletti, G. and co-authors (2023) ``Harnessing AI and Computing to Advance Climate Modeling and Prediction.'' \textit{Nature Climate Change} 13: 887 -- 889
\item
Scrucca, L. (2013) ``GA: A Package for Genetic Algorithms in R.'' \textit{Journal of Statistical Software} 53(40): 1 -- 37.
\item
 Susskind, J., Blaisdell, J., Iredell, L., Lee, J., Milstein, A., Barnet, C., Fishbein, E., Manning, E., and Strow, L. (2020) ``Algorithm Theoretical Basis Document: AIRS-Team Retrieval For Core Products and Geophysical Parameters: Versions 6 and 7 Level 2.'' Jet Propulsion Laboratory.
 \item
Tennant, P.W.G., Arnold, K.F., Ellison, G.T.H., and Gilthorpe, M.S. (2022) Analyses of ‘Change Scores’ Do Not Estimate Causal Effects in Observational Data''. \textit{ International Journal of Epidemiology} 51(5): 1604 -- 1615. https://doi.org/10.1093/ije/dyab050
\item
Therneau, T.M., and Atkinson,E.J. (2023) ``An Introduction to Recursive Partitioning Using the RPART Routines.''  \textit{Mayo Foundation} 12/5/2023.
\item
Thrastarson, H. (2021) ``AIRS/AMSU/HSB Version Level 2 Product User Guide.'' jet Propulsion Laboratory
\item
Tian, B., Manning, E.,  Roman, J., Thrastarson, H., Fetzer, E., and Monarez, R. (2020) ``AIRS Version 7 Level 3 Product User Guide'', Jet Propulsion Laboratory.
\item
Touchette, H. (2009) ``The Large Deviation Approach to Statistical Mechanics.'' arXiv: 0804.0327v2 [cond-math.stat-mech]
\item
Garth, H.H. and Mills, C. (1958) \textit {From Max Weber: Essays in Sociology.} Oxford University Press.
\item
Waldman, D.M. (2000) ``Discrete Choice Models with Choice-Based Samples.'' \textit{The American Statistician} 54(4): 303 -- 306.
\item
Wang, S., Dai, Y., Shen, Y., and Xuan J. (2021) ``Research on Expansion and Classification of Imbalanced Data Based on SMOTE Algorithm.'' \textit{Scientific Reports} 11(24039) https://doi.org/10.1038/s41598-021-03430-5.
\item
Zhang, L., Risser, M.D., Wehner, M.F., and O'Brien, T.A. (2024) ``Leveraging Extremal Dependence to Better Characterize the 2021 Pacific Northwest Heatwave.'' \textit{Journal of Agricultural, Biological, and Environmental Statistics.}  https://doi.org/10.1007/s13253-024-00636-8.
\item
Zurita-Gotor, P., and Vallis, G.K. (2011) ``Dynamics of Midlatitude Tropopause Height in an Idealized Model.'' \textit{Journal of Atmospheric Sciences} 68(4): 823 -- 838.
 
\end{description}

\bmhead{Acknowledgements}
Eric Tchetgen Tchetgen offered very insightful concerns about endogenous sampling.
\end{document}